\documentstyle[preprint,aps,prb]{revtex}
\begin{document}

\newcommand{\scriptbox}[1]{\mbox{\scriptsize #1}}
\newcommand{\displayfrac}[2]{\frac{\displaystyle #1}{\displaystyle #2}}


\draft
\tighten
\title{Superconducting Order Parameter Symmetry in Multi-layer
Cuprates}
\author{J. Maly, D. Z. Liu and K. Levin}
\address{Department of Physics, University of Chicago, Chicago,
Illinois, 60637}
\address{\rm (Submitted to Physical Review B on August 24, 1995)}
\maketitle

\begin{abstract}
In this paper we
classify the allowed order parameter symmetries in multi-layer cuprates
and discuss their physical consequences, with emphasis
on Josephson tunneling and impurity scattering.  Our solutions
to the gap equation are based on
highly non-specific forms for the inter- and intra-plane pairing
interactions in order to arrive at the most general conclusions.
Within this framework, the bi-layer ($N=2$) case is
discussed in detail with reference to YBCO and BSCCO and the related
Landau-Ginzburg free energy functional. Particular attention is paid to the
role of small orthorhombic distortions as would derive from
the chains in YBCO  and from superlattice effects in BSCCO,  which give
rise to a rich and complex behavior of the multi-layer order parameter.
This order parameter has $N$ components associated with each of the $N$
bands or layers.  Moreover, these components have specific
phase relationships.  In
the orthorhombic bi-layer case the $(s,-s)$ state is of special interest,
since for a wide range of phase space, this state exhibits $\pi$ phase
shifts  in corner Josephson junction experiments. In addition, its transition
temperature is found to be insensitive to non-magnetic inter-plane
disorder, as would be present  at the rare earth site in YBCO, for example.
Of particular interest, also,  are the role of van Hove
singularities which are
seen to stabilize states with $d_{x^2 - y^2}$-like symmetry, (as well as
nodeless $s$-states) and to elongate the gap functions along the four
van Hove points, thereby leading to a substantial region of
gaplessness.  We find that $d_{x^2 - y^2}$-like  states are general solutions
for repulsive interactions; they possess the fewest number of
nodes and therefore the highest transition temperatures. In this way,
they should not be specifically associated
with a spin fluctuation driven pairing mechanism.
\end{abstract}

\makeatletter
\global\@specialpagefalse
\def\@oddhead{REV\TeX{} 3.0\hfill Levin Group Preprint, 1995}
\let\@evenhead\@oddhead
\makeatother

\section{Introduction}
\label{Intro}

The question of the order parameter symmetry in layered
cuprate superconductors is one of the most important
issues currently under debate.  Despite strong evidence for
a $d_{x^2-y^2}$  state,\cite{ScalRev} at least in one particular cuprate
(YBCO), there are still experimental inconsistencies.\cite{DynesRev,LevinRev}
This
matter is complicated further by the  complexities of YBCO
associated with the double copper-oxide plane structure.
The same bi-layer unit is shared with the bismuth 2212
compounds (BSCCO), where recent
photoemission data,\cite{Shen,NormCamp-up95}
seem also to favor a $d_{x^2-y^2}$
symmetry. In the former compound,\cite{Camp-prl64}
if not in the latter,\cite{DingBellCamp} there
is evidence that the double layer unit leads to two bands
each of which crosses the Fermi surface and each of which
presumably has a distinct superconducting order
parameter. It is not a priori clear whether these two gaps
have the same or opposite phase, nor whether they  are
predominantly associated with the same or different
irreducible representations of the tetragonal symmetry
group. It is, therefore, essential to provide a systematic
classification of the order parameter in these bi-layer (and
more general $N$-layer)  systems before meaningful and
unambiguous inferences can be deduced from the
experimental data.

It is the purpose of this paper to expand upon earlier work
by ourselves \cite{us} and other groups
\cite{LiechtMazAnd,KlemmLiu,BulScal,BulZys,KetEf}
by classifying and
establishing physical consequences of  the various order
parameter symmetries  in $N$-layer systems.  Of particular
interest are those parameter sets which lead to $\pi$
junction behavior \cite{Tsuei,BrawnOtt,WollvanHarl,MathaiGim}
in an $a,b$-plane Josephson configuration. It
is widely assumed that the observation of these $\pi$ phase
shifts  provides the strongest evidence yet for the $d_{x^2-
y^2}$ state in YBCO. By contrast, we find that in the more
general orthorhombic bi-layer case, in part because of van
Hove effects, $\pi$ phase shifts are fairly widespread and
not uniquely associated with the $d_{x^2-y^2}$ state.
Moreover, the proximity to the van Hove singularities leads
to a stabilization of the $d_{x^2-y^2}$ symmetry  for a variety of
different models for the pairing interaction, beyond the
simple spin fluctuation model.\cite{TsueiNewns}
It also is associated with a
considerable distortion of the gap function away from the
ideal representation of the $d_{x^2-y^2}$ state. In order to
understand the physical consequences of various order
parameter sets we investigate the role of intra- and
inter-layer impurities, include  a discussion of Josephson coupling,
and present a more general analysis of our results in the
context of both a Landau-Ginzburg theory and the solution
of the arbitrary $N$-layer problem.

A number of authors have considered the possibility of
electron tunneling and pair interactions of electrons
on different planes in the context of a bilayer structure.
Bulut and Scalapino \cite{BulScal} used numerical solutions of
a strong coupling model to show that when both
interlayer interactions and hopping were included in a
bilayer model two
competing states arose, one with $d_{x^2-y^2}$ solutions
in phase on both sub-bands of the Fermi-surface and the
other with $s$-states of opposite sign.   Similar observations
were
made by Liechtenstein and co-workers \cite{LiechtMazAnd}
who argued that the
more
probable situation for spin fluctuation induced superconductivity,
corresponded to a  pair of out-of-phase $s$ states.
It should be noted that throughout this paper we will use the
generic notation "$s\,$"  ( and "$d\,$" ) as
applying to gap functions which have the same (or different) signs
under a rotation of the wave-vector by $\pi / 2 $. It is important to
be particularly clear on our  notational convention since we
emphacize the role of orthorhombicity in our work. Among other
things, this orthorhombicity leads to what we  refer to as "$s$-$d\,$"
or
"mixed states",  (although the overall sign under a $\pi / 2 $ rotation
remains either $+$ or $-$,
depending on which component is dominant).
The possibility of mixed  $s$-$d$-states
has been discussed within Landau-Ginzburg theory, in  a  mono-layer
\cite{LiKoltJoynt} as well as a  bi-layer context.\cite{KobukiLee}
In the latter, a spontaneous $s$-$d$ mixing was
considered,  using  microscopic arguments based on an RVB
decoupling scheme.  While most bi-layer studies considered
tetragonal
symmetry, earlier work by our own group,\cite{us} using the more
conventional BCS pairing formulation,  investigated the role
of
a small amount of $a,b$-axis anisotropy in the bi-layer
problem. This small orthorhombicity appears to be amplified
by van Hove effects. As a consequence, the solution
corresponding to a pair of out of phase
$s$-wave  states frequently has the interesting and
important physical consequence of  leading to $\pi$
phase shifts in a corner Josephson tunneling
experiment.\cite{Tsuei,BrawnOtt,WollvanHarl,MathaiGim}
It should be stressed that orthorhombicity is assumed to enter via
the chains in YBCO.  There are alternate scenarios for this material
which explicitly build in the chain bands
\cite{MazGolZai,Pokrovski,Combescot}
and their contribution to the superconducting gap.  Here we ignore
these explicit
effects, in large part because recent tunneling studies indicate that
the order parameter behavior is largely unaffected by reducing the
oxygen stoichiometry to the limit where the chains are strongly
fragmented \cite{BrawnOtt}
(and thus presumably irrelevant to the superconductivity).

The more general $N$-layer problem has been studied by
Bulaevski and Zyskin \cite{BulZys}
as well as  Klemm and Liu.\cite{KlemmLiu}
In these calculations, it was also assumed that
the $c$-axis consisted of a coherently coupled stack of
bi-layer structural units.
Here we presume that the
considerable evidence for the absence of $c$-axis coherence
\cite{CooperGray,PWAnd-prl67}
requires a different starting point. While, for simplicity, we
focus on an isolated $N$-layer complex, it is clear that higher
order effects associated with incoherent coupling between
unit cells must ultimately be included. Such incoherent
coupling can be introduced  following, for example,
Ref.~\onlinecite{RojoLevin}. An additional complexity was raised in  even
earlier studies by
Efetov and Larkin \cite{EfLar} who pointed out that a
complete treatment of interlayer electron
pairing required the  consideration of a triplet pairing
state odd under exchange of layer indices. This case was
further investigated by Kettemann and Efetov \cite{KetEf}
as well as Klemm and Liu \cite{KlemmLiu} who argued
that the mixing of triplet and singlet states would generally
lead to a second transition below $T_c$.  Thus far there is no
evidence for this second transition, nor is there much
support for the triplet pairing state.

There are important issues of controversy implicit in the
work discussed here.  There appears to be little doubt that
the components of the bi-layer couple magnetically.
However, it has not been persuasively demonstrated that
there is a coherent electronic coupling $t_{\perp}$  within
the unit cell.\cite{ChakraAnd}
Neutron experiments \cite{RosMign,Tranq,Mook} on
YBCO provide evidence for a $c$-axis modulation associated
with the spacing of the planes within the bi-layer. While this
can be explained by magnetic coupling alone,\cite{ZhaSiLevin,LiuZhaLevin}
earlier photoemission studies on this cuprate
have reported the observation of two copper oxide plane
bands.\cite{Camp-prl64} The situation in BSCCO  is even
less certain with two photoemission groups reaching
opposite conclusions.\cite{Shen,DingBellCamp}
For the purposes of the present
paper it will be assumed that there is coherent coupling
between the layers, though not between the unit cells. This
assumption is based in large part on the demonstrated
intra-bi-layer magnetic interactions which suggest a
moderate degree of communication between the layers.
Moreover, on this basis it may be presumed that there are
direct or indirect electronic interactions within the bi-layer
complex which must be included in any complete theory of
the superconductivity.

Similarly controversial is the origin of a possible $d_{x^2-
y^2}$ state. While this state is consistent with spin
fluctuation mediated superconductivity,\cite{ScalRev,PinesRev}
in this paper we
demonstrate that it appears more generally as a solution to
the gap equation in the presence of repulsive interactions.
Providing this repulsion has some wave-vector dependence
(near, although not necessarily pinned at the anti-ferromagnetic
position $(\pi, \pi)$),  we find the $d_{x^2-y^2}$
state to be stable.  Thus, experimental observation of this
symmetry appears to provide more support for
superconductivity arising from repulsive interactions than
for any detailed superconducting mechanism.
While there are problems associated with interpreting
various experiments within a $d_{x^2-y^2}$ context, among
the most perplexing from a theoretical viewpoint
\cite{Millis,RadLevin-dirt} is the evident impurity insensitivity of
the superconducting transition temperature.\cite{Sun-prb50}
Moreover,  substitution at the rare earth site in YBCO with
both magnetic and non-magnetic atoms,  between the
planes of the bi-layer,  leads to no variation in $T_c$, except
in the special case of Pr.  Here we address this
fastinating puzzle and demonstrate that for the case of two
out of phase nodeless $s$-states in an isolated bi-layer
configuration, Anderson's theorem applies to
inter-plane substitutions. This provides additional
motivation for studies of this particular pairing state.

An outline of the paper is as follows.
In Section~\ref{bilayer} we focus on the bilayer problem as a
prototype
of the $N$-layer system. We introduce a generalized model for the
pairing interaction (Section~\ref{model}) and then
point out some general properties of
solutions which arise in these systems (Section~\ref{gap-symmetry}).
In Section~\ref{tunnel} we examine the
consequences of the bilayer structure on
Josephson tunneling experiments between cuprates and
conventional superconductors
as well as the problem of twinned samples.
We then construct (Section~\ref{LG}) the
free energy functional in the
Landau-Ginzburg limit, addressing the effects of interlayer
tunneling and  orthorhombicity.
Impurity scattering will be discussed in Section~\ref{dirt}, where
we investigate the circumstances under which Anderson's
theorem holds for the case of two out of phase $s$-states.

In Section~\ref{Analysis} we turn to the more technical issues of a
general weak coupling
treatment of the problem of $N$ two dimensional layers. The
reader uninterested in the more mathematical details can
skip directly to the conclusions in Section~\ref{conclusion}.
The generalized gap equation in the $N$ layer problem is
derived (Section~\ref{Calc-Gap}) and the two competing
states are discussed in Section~\ref{Ideal}.
These correspond to natural analogues of the in-phase and
out-of-phase states of the bi-layer system.  We conclude
with a general discussion (Section~\ref{van-Hove})
of the role of van Hove effects.
Two states are found to
take maximal advantage of the van Hove singularities: these are
the $d_{x^2 - y^2 }$ and nodeless $s$-states.
Some of the technical aspects of these calculations as well as a complete
solution of the bilayer problem are relegated to
Appendices~\ref{show-N-layer} and \ref{IB-pairing} respectively.

Our conclusions are presented in Section~\ref{conclusion}.

\section{The $N=2$ Case}
\label{bilayer}

\subsection{Gap Equation and Model Interaction}
\label{model}

We initially study the simpler bi-layer system  ($N=2$) in
order to
establish notation and to discuss physical results in a more
familiar context. Although only singlet intraband pairing is
considered in
the main part of the text,
a more complete weak coupling calculation
of the bilayer problem will be presented in Appendix~\ref{IB-pairing}.
Many of the interesting
features of bilayers arise in the general $N$-layer
problem with only slight modification. However, this more
general case
will be discussed in detail  in Section~\ref{Analysis}.
Section~\ref{gap-symmetry} focuses
primarily on the linearized gap equation. Throughout
Section~\ref{bilayer},
particular attention will be devoted to elucidating the role of
orthorhombicity.

The gap equations for two copper oxides planes are given by
\begin{mathletters}
\label{bilayer-gap}
\begin{eqnarray}
\Delta_+ + \Delta_- & = &
-\sum_{\bf q}V_{\parallel}
\left( \Delta_+
\frac{\tanh\frac{1}{2}\beta E_{+}}{2E_{+}}
+ \Delta_-\frac{\tanh\frac{1}{2}\beta E_{-}}
{2E_{-}}\right)   \\
\Delta_+ - \Delta_- & = &
-\sum_{\bf q}V_{\perp}
\left( \Delta_+
\frac{\tanh\frac{1}{2}\beta E_{+}}{2E_{+}}
- \Delta_-\frac{\tanh\frac{1}{2}\beta E_{-}}
{2E_{-}}\right)
\end{eqnarray}
\end{mathletters}
where we define
\begin{equation}
E_{\pm}^2 = \epsilon_{\pm}^2 + \left|\Delta_{\pm}\right|^2
\end{equation}
This equation was previously discussed in Ref. \onlinecite{us}. Its
general
derivation is presented in the $N$-layer context in
Section~\ref{Calc-Gap}.
The two order parameter
components $\Delta_+$ and $\Delta_-$ refer to
pairing on the bonding and anti-bonding bands of the
Fermi surface respectively.
Throughout this paper we will transform
between the band and layer indices. The two gap parameters
may be related to order parameters in the layer language
by the equations
\begin{equation}
\begin{array}{rcl}
\Delta_+ & = & \Delta_{\parallel} + \Delta_{\perp}  \\
\Delta_- & = & \Delta_{\parallel} - \Delta_{\perp}
\end{array}
\label{band-to-layer}
\end{equation}
where $\Delta_{\parallel}$ refers to singlet pairing of
electrons
within individual layers and $\Delta_{\perp}$ refers to
singlet pairing
of electrons on different layers.
Finally,  the normal state energy dispersion is given by
$\epsilon_\pm = \xi \mp t_{\perp}$,
where $t_{\perp}$ is the interlayer hopping matrix element
and $\xi({\bf q})$ is the single particle dispersion
within the copper oxide planes.
It should be noted that allowing  the interlayer
tunneling to depend on ${\bf q}$ in the manner suggested by
Chakravarty and Anderson \cite{ChakraAnd} as well as
Andersen et al. \cite{Andersen}
\begin{equation}
\frac{t_{\perp}}{4}\left( \cos q_x - \cos q_y \right)^2
\label{t-perp-q}
\end{equation}
has no qualitative effect on the results of this paper. This
arises because for the model of Ref. \onlinecite{ChakraAnd}
the inter-plane tunneling deviates most from the constant value
$t_{\perp}$ along the diagonals
of the Brillouin zone (where Eq.~(\ref{t-perp-q}) vanishes).  It is in
precisely this region where the density of states
contributions are
smallest, since, as will be discussed in more detail below,
they lie away from the four van Hove points.

As noted above,
the bilayer system admits two other pairing states, an
interlayer
triplet state $\Delta_{\perp}^t$ which is odd under
interchange
of layer labels and a second intralayer singlet
state $\Delta_{\parallel}^o$
where the order parameter has opposite phase on each
of the two layers.
Both of these  correspond to
inter-band pairing, i.e. pairing
between  electrons on different sub-bands of the Fermi
surface. Thus they appear as off-diagonal elements of
$\Delta$
in the band representation. Because this inter-band
pairing is associated with a second phase transition (when
$t_{\perp}$ is finite) as
well as a triplet state, it appears to be currently of less
physical interest than the  singlet, intra-band pairing which
we consider here. For the purposes of completeness,  these
more exotic states are considered  in
Appendix~\ref{IB-pairing}.

While our analysis of Eq.~(\ref{bilayer-gap}) is expected to be
generally valid,  we present here a weak coupling
treatment,  in large part because our emphasis is on the
symmetry of the order parameter, rather than on the
detailed values of the transition temperatures.
Furthermore we adopt a  model interaction which is
believed to be sufficiently generic so as to encompass most
pairing mechanisms in the literature. Thus it includes  the
well studied spin fluctuation interaction which is strongly
peaked
at quasi-momentum transfers of ${\bf Q}=(\pi, \pi)$,\cite{ScalRev}
as well as phonon based mechanisms.\cite{Abrikosov}
Alternative scenarios
\cite{Varma} may be associated with intermediate ${\bf Q}$
values.
In the present model,  the in plane and inter-plane pairing
interactions are given by
\begin{equation}
\begin{array}{rcl}
V_{\parallel}({\bf q},{\bf q'}) & = & \lambda_{\parallel}\,
\chi_{{\bf Q}_0}({\bf q-q'})  \\
V_{\perp}({\bf q},{\bf q'}) & = & \lambda_{\perp} \,
\chi_{{\bf Q}_0}({\bf q-q'})
\end{array}
\label{V-lambda-chi}
\end{equation}
where we define the generalized susceptibility, which depends on
the momentum transfer
\begin{equation}
\chi_{{\bf Q}_0}({\bf q}) = \frac{1}{h}\sum_{{\bf Q}=R{\bf
Q}_0}
\frac{1}{\left[ 1-J_0\left(\cos(q_x-Q_x)+\cos(q_y-
Q_y)\right)\right]^2}
\label{chi}
\end{equation}
This model incorporates several parametrizations, allowing
the variation of
the peak location through ${\bf Q}_0$, the peak strength via
$J_0$, and the coupling strength  and sign through
$\lambda$. We consider $J_0$ to be below the critical value
$0.5$. The
interaction contains the underlying symmetry of the lattice.
This is imposed  by summing over all elements $R$ of
the point group, $h$ being the order of the group. Thus
when
${\bf Q}_0$ does not lie on a symmetry element of the lattice
then the interaction will have eight peaks in a tetragonal
$D_{4h}$ lattice and four peaks in an orthorhombic lattice.

Our aim is to incorporate some of the key features of the
known band structure in various cuprates. Thus the lattice
point group will be assumed either to be
$D_{4h}$ in the case of LSCO or $D_{2h}$ in the case
of YBCO and BSCCO. We investigate two different
kinds of broken tetragonal symmetry  in which either
principle axes are retained as symmetry planes
(orthorhombic) in the
case of YBCO or the diagonals are retained as planes
of symmetry (rhombohedral)\cite{fn-rhombo} in the case of BSCCO.
\cite{Fukushima}
The band structures which we use throughout this paper
are taken from Ref. \onlinecite{RadNorm}.
In plane energy dispersions of the normal state are given
by an expansion
of the form
\begin{equation}
\xi({\bf q}) = \sum_{i=0}^7 t_i \eta_i({\bf q})
\label{xi-def}
\end{equation}
where the basis functions $\eta_i$ are listed in
Table~\ref{bs-tab}.
Also indicated in the table are  the
various band structures
we will be considering.
It is important to note that the degree to which the tetragonal
($D_{4h}$)
symmetry is broken is reflected in  the last two parameters of the
table which we call $t_6$
and $t_7$. The size of these orthorhombic contributions was chosen,
for illustrative purposes,  to be small, but otherwise arbitrary.

\subsection{Order Parameter Symmetry in Bilayers}
\label{gap-symmetry}

\subsubsection{Properties of Mono-layer Solutions}
\label{1l-solutions}

Solutions to the single layer
problem provide intuition into those of the $N$-layer case,
particularly when the
interlayer hopping is relatively small.  It is useful,
therefore, to characterize the allowed order parameters in
the one layer limit for each of the three bandstructures
discussed above, before moving to the bi-layer case.
In the one  layer system the functional form of the
superconducting
order parameter depends only on the sign and the shape (as
distinct from the magnitude)  of the pair interaction and on
the electronic structure.
To illustrate these pair interaction and electronic
effects,  we vary the position of
the peak location ${\bf Q}$ and plot the associated form of the
order parameter in this two dimensional space, for each of
three model bandstructures.  The interaction is assumed
repulsive ($\lambda>0$)
in this and subsequent phase plots; attractive
interactions in general give rise to nodeless $s$-wave
solutions for any value of ${\bf Q}$.
The magnitude of the interaction plays no role in determining the
symmetry of the order parameter in the single layer problem,
it  results only in a variation of the transition temperature.
By contrast, in the two layer case
the (relative) magnitudes of the in plane and out of plane
interactions enter via Eq.~(\ref{bilayer-gap}) in an important way.
We will examine this effect subsequently.

Other parameters of our model interaction have a lesser
influence on the form and regions of stability of
the various solutions.
Variation of the peak width $J_0$
affects $T_c$ to a greater extent than it does the symmetry
of the actual solution.
The component of the interaction which most influences the
symmetry of the superconducting order parameter is
the degree to which the vector ${\bf Q}$ connects positive
and negative lobes of the pair wave function on the Fermi
surface.\cite{PinesRev} As will be illustrated under more
general circumstances below,  near half filling, when
only nearest neighbour hopping is included, the points at
which the $d_{x^2-y^2}$ state has maxima are separated by
${\bf Q}=(\pi,\pi)$, while the extrema of $d_{xy}$ are
separated by $(\pi,0)$ and the eight lobed $s_{xy}$ state
has positive and negative lobes separated by
$(\pi/2,\pi/2)$.
These are precisely the points in the space of ${\bf Q}$
near which these various solutions are favoured.

Figure~\ref{phase-LSCO-fig}
represents this
"phase diagram" for the LSCO case. It is a
map of the irreducible representations
of the tetragonal, $D_{4h}$ lattice which correspond to the
solutions
with the highest $T_c$.
We stress that the simple functional forms of the order parameter,
such as are implicit in the notation "$d_{x^2-y^2}$" etc.,  can
be misleading. The gap  parameters may involve important
contributions from higher order terms in the associated
representation.  Moreover when  tetragonal symmetry is
broken, states may be highly admixed with states of
other representations.
We will use this notation with care, applying it
only to certain solutions of systems with tetragonal symmetry.
When necessary we will avoid any ambiguity by using the
common group theoretic nomenclature for the various
irreducible representations.\cite{Sahu}

Because proximity to the van Hove singularity is found to
play such a crucial role, we present two limiting cases
corresponding  to choices for the Fermi energy near
(\ref{phase-LSCO-fig}a)
and far from (\ref{phase-LSCO-fig}b) the van Hove singularity. The
order parameters shown plot the actual solutions for
${\bf Q}=(\pi,\pi)$, at which the $B_{1g}$  ( "$d_{x^2-y^2}$" )
solution is found, at ${\bf Q}=(\pi /2,\pi / 2)$, where an eight-lobed
$A_{1g}$ solution, the so called
"$s_{xy}$"  state, exists and at  ${\bf Q}=(\pi,0)$, where the
$B_{2g}$ "$d_{xy}$" state is most stable. Moreover, these
symmetries  persist over an extended region of
${\bf Q}$-space as outlined by the solid lines in the figures. States with
increasing numbers of nodal planes, including
states in the $A_{2g}$ irreducible representation, appear
as ${\bf Q} \rightarrow (0,0)$.  Indeed, because this region is
so complicated (in the case of repulsive interactions),  no
detailed solutions are indicated. The region where the "$d_{x^2-y^2}$"-like
solution persists is shaded in this and subsequent figures.

Figures~(\ref{phase-LSCO-fig}a) and~(\ref{phase-LSCO-fig}b)
demonstrate that proximity to the van
Hove singularity leads to a modestly extended region with
$B_{1g}$ or "$d_{x^2-y^2}$" symmetry. Moreover as
the Fermi energy approaches the singularity,  the lobes on
the "$d_{x^2-y^2}$" state become sharper.  In this way the
order parameter takes full advantage of the four van Hove
points which coincide with each of the four maxima in this
gap function. Since the gap function is sharply peaked in the direction
of the principle axes it is very small in a large region about the
nodal lines along the diagonals of the Brillouin zone. A large
gapless region will have consequences on the thermodynamic
properties of the samples at low temperatures and may have been
observed in ARPES measurements on BSCCO.\cite{Shen,NormCamp-up95}

Breaking of the tetragonal symmetry, as in the case of the
orthorhombic lattice of
YBCO, serves to further enhance these van Hove effects.
This is shown in  Figure~\ref{phase-YBCO-fig}
where, as in the LSCO case
above,  the gap equation solutions are plotted at the special
symmetry points and the lines indicate the boundaries
of the region over which the representative solution
persists. Solid lines separate states belonging to different
irreducible representations and dotted lines separate
states of different symmetry within a representation.
Figures~(\ref{phase-YBCO-fig}a)  and~(\ref{phase-YBCO-fig}b)
demonstrate clearly that the
phase space occupied by the "$d_{x^2-y^2}$" solution  is
dramatically increased in the vicinity of the van Hove point.
Furthermore, the lobes of this gap function again become  more
extended along the principle axes,
compared with similar solutions farther from the van Hove
point.

Finally, the results in the BSCCO case are presented
in Figure~\ref{phase-BSCCO-fig}.
We see here that, contrary to the two previous cases,
the phase space in which the gap has the same sign
along the $a$ and $b$ directions (i.e, the totally symmetric
irreducible representation  in  the $D_{2h}$ lattice)  occupies
a somewhat larger region of parameter
space when the Fermi level is near the van Hove point.
This is the consequence of a stabilizing admixture
of an isotropic $s$-wave component.
On the other hand, states with an $a,b$-axis $\pi$ phase
shift dominate the phase space away from the van Hove points.

The above three figures lead to an important conclusion: the
"$d_{x^2-y^2}$" solution exists in a large region of
parameter
space away from ${\bf Q}_{\scriptbox{AF}}$. Therefore this state
should be considered as an obvious candidate form for
the order parameter in any mechanism involving pairing
via a predominantly repulsive interaction. Stated
alternatively, observation of this form for the gap in an
experiment
is in itself not proof of any specific pairing mechanism.
Furthermore, it also appears from these figures that,  quite
frequently, a
single mechanism can give rise to widely varying forms of
order parameters depending upon the details of the
band structure, doping level, lattice symmetry
and other subtle features of the material.

\subsubsection{Bi-layer Effects}
\label{2l-solutions}

In bi-layer materials there are two possible states which
compete as solutions to the gap equation.  These may be
crudely classified by referring to the states as "inter-plane
and intra-plane dominated solutions". These two competing
sets of solutions were discussed in a previous short
communication by our group,\cite{us} as well elsewhere.
\cite{LiechtMazAnd,BulScal,KobukiLee,UbbensLee}
By setting $t_{\perp}=0$ in the gap
Eq.~(\ref{bilayer-gap}) we see that
$\Delta_{\parallel}$ and $\Delta_{\perp}$
uncouple from each other at $T_c$. Superconductivity
can thus be either due to interlayer pairing or due to
intralayer pairing depending on the relative
magnitudes and signs of $\lambda_{\parallel}$ and
$\lambda_{\perp}$.
In the band representation in-plane pairing
yields a solution which has the same
phase on both bands while the interlayer pairing
solution undergoes a $\pi$ change in phase from
one band to the other. This sign change follows directly from
Eq.~(\ref{band-to-layer})
in the extreme limits in which one or the other order
parameter (in the layer basis)  is the larger.

It is important to stress that we will refer to these two
competing solutions throughout this paper. They occur in bi-layer as
well as $N$-layer systems. In all cases where $t_{\perp}$ is finite, one
solution
is stable and the other metastable.  Thus, although one may refer to
their
associated transition temperatures, only the larger of the two has
any
physical meaning. As a consequence, there is nothing to prevent
these
two competing states from having a different order parameter
symmetry.

As in the one layer case, repulsive interactions are found to
lead to nodal states such as the $d$-states or multi-lobed
$s$-states discussed above,  while attractive interactions
require nodeless states of $s$ symmetry. In the presence of
two different types (signs) of in plane and out of plane
interactions, these effects provide a rich and complex
structure for the competing order parameter states. Under
these conditions and when there is no coherent inter-plane
hopping, the two solutions may belong to different
symmetry groups. When  $t_{\perp}$  is finite, the solutions
couple while still preserving the underlying lattice
symmetry.
Thus mixing of $s$- and $d$-states within the linearized
gap  requires both orthorhombicity and inter-plane hopping.
More general $s$-$d$ mixing  effects  will  occur at
higher
(quartic) order in a Landau-Ginzburg expansion, and will be
discussed in more detail in a subsequent section.

To illustrate the behavior of the bi-layer order parameter
and the two competitive states, we consider for concreteness
the case of an attractive
interplane interaction and
a repulsive intraplane interaction peaked at ${\bf Q} =
(\pi,\pi)$. This model may be viewed as derived from a
generalized spin fluctuation model, where the inter-layer
or $c$-axis spin susceptibility is included as a perpendicular
component of the pairing interaction. This case was
previously discussed by a number of different groups.
\cite{us,LiechtMazAnd,BulScal,UbbensLee,Normand}
While we choose
this peak location simply for illustrative reasons, our results
can easily be extended to other values
of ${\bf Q}$.
Under these circumstances, the order parameter in the
band picture
has the schematic form
\begin{equation}
\left(\begin{array}{c} \Delta_{+} \\ \Delta_{-}\end{array}\right) =
\left(\begin{array}{c} s_{\perp} + d_{\parallel} \\
-s_{\perp} + d_{\parallel}
			\end{array}\right)
\label{s,d-mix-band}
\end{equation}
If the lattice is tetragonal then either $s_{\perp}$ or
$d_{\parallel}$ will vanish. Thus when $\lambda_{\perp}$ is
zero the solution will have $d$-symmetry.  Moreover, both
band gaps are in phase.  Similarly when $\lambda_{\parallel}$
vanishes the symmetry is $s$-like and both gaps are out of
phase. The transition from one
type of solution to the other is governed by size of
$\lambda_{\perp}/\lambda_{\parallel}$.  Frequently the
cross over from the in-plane to the inter-plane dominated
regime may occur despite the fact that the ratio
$\lambda_{\perp}/\lambda_{\parallel}$ is small
compared to one. What determines this cross-over is the
nature and sign of the competing interactions and
the bandstructure.

If tetragonal symmetry
is broken even more complicated situations obtain. Both
components can be nonzero
simultaneously and the transition is a smooth
one as a function of increasing $\lambda_{\perp
}/\lambda_{\parallel}$. The regions where the various
types of solution
exist in YBCO are illustrated in Figure \ref{bilayer-fig}
for
the example discussed above.  Plotted on the vertical axis is
the fractional hole doping $x$. We estimate the
physical values of this parameter, for a range of different
oxygen stoichiometries in YBCO to vary from 0.1 to 0.3.
The figure illustrates the parameter regimes where the dominant
component of the order parameter on the two bands is
$(d,d), (s,d)$ and $(s,-s)$.\cite{fn-sd-mix}
The region below the solid line involves states in which
the sum of the pair wave functions on the two bands
changes sign under $\pi / 2$ rotation of the axes and can
thus yield $\pi$ phase shifts in $a,b$-axis Josephson
tunneling experiments.\cite{Tsuei,BrawnOtt,WollvanHarl,MathaiGim}
Moreover, the shaded regions denote
states in which the sum of the two order parameters changes
sign even though one or both are nodeless.
This issue will be addressed in detail in Section \ref{tunnel}.
The inset plots the density of states as a
function of energy. There are actually four van
Hove points associated with orthorhombic splitting of the
two bands.
For clarity the doping level at which the first of
these occurs is plotted in the
main portion of the figure as a dotted line. Note that the
region labelled by
$(s,-s)$ occupies the largest fraction of phase
space when the band is near the van Hove points
in general and in particular
when one sub-band of the Fermi level is closed about
the $\Gamma$ point of the reciprocal lattice
and the other about $X$.

The evolution of these solutions  with increasing
$\lambda_{\perp }/\lambda_{\parallel }$ is illustrated in
Figure \ref{s,d-mix-fig}.
Indicated here is the calculated
shape of the gap functions for the two bands of the bi-layer.
Their relative phase is also noted.  It can be seen that the
pair wave function on each band has a greater amplitude
along one symmetry axis than along the other, and the axis
along which the pair wave function has its maximum
differs from one sub-band to the other. These results are
plotted for very small orthorhombicity and for Fermi
energies close to the van Hove point.  On the
basis of the latter assumption, the small orthorhombic
effects are greatly magnified. The figure labels (a)-(d)
correspond to (a) two in phase $d$-wave dominated states,
(b) one band nodeless and the other nodal, (c)  two out of
phase nodeless states which show greatly elongated lobes,  and (d)
to two out of phase $s$-wave dominated states which are more isotropic.
The relative phase space which these solutions occupy can be
seen from the previous figure.

Moreover, the detailed shape of the order parameter has
important physical consequences.  Thus for case (c), for
example, the solutions are in the inter-plane regime where
$\Delta_{\perp}$ is dominant.  Here, however,  the order
parameter
is nodeless and yet the combination of the two gaps has
features of a $d$-wave solution in $a,b$-axis
as well as $c$-axis Josephson
tunneling experiments.\cite{Sun-prl72}
Case (b) is associated with one $s$-like
and one $d$-like state. It will exhibit both power law behavior in
thermodynamical properties, as well as the $\pi$ phase
shifts in corner junction experiments, which are also seen in cases (a)
and (c).  These Josephson experiments and their relation to the
order parameter symmetry will be discussed
in more detail in the next section.

\subsection{Josephson Tunneling: Phase Coherence Across Domain Boundaries}
\label{tunnel}

Josephson tunneling experiments have been key to elucidating the
order
parameter symmetry in the YBCO cuprate family.
\cite{Tsuei,BrawnOtt,WollvanHarl,MathaiGim,Sun-prl72}
Within this class, two types of
measurements have been performed. These involve   $a,b$- and $c$-axis
tunnel
junctions.  In the former category SQUID geometries have been
investigated.\cite{BrawnOtt,WollvanHarl,MathaiGim}
These consist of two junctions with Pb counterelectrodes,  whose
interference
pattern  yields information about the relative phase of the order
parameter along
various axes in the $a,b$-plane.  In "corner junctions",
$\pi$ phase shifts of the order parameter between the $a$ and $b$ axes
have been
observed and attributed to the sign change upon a $\pi / 2 $ rotation
of the wave
vector, within a
$d_{x^2-y^2}$ state.  This interpretation is further confirmed by
"edge" junctions
which probe the interference along a single face of the material; here
no phase
shifts are found to be present. These conclusions are the same for
both twinned
and untwinned crystals.  A variant on this geometry are the ring
experiments \cite{Tsuei}
consisting of  YBCO segments with different grain boundary
orientations.
Observation of a non-zero, half integer spontaneous flux threading
the ring,
for specific orientations of the grain boundaries, again provides
support for a
$d_{x^2-y^2}$ state.

An alternate Josephson tunneling geometry has been investigated by
Sun et al.\cite{Sun-prl72}
These authors observe finite $c$-axis Josephson currents between a Pb
counterelectrode and a YBCO sample. For the simplest representation
of the
$d_{x^2-y^2}$ symmetry, no Josephson current should be present.
While it is
possible to invoke orthorhombicity \cite{ODonCarb} to explain these
non-
vanishing results,  this hypothesis seems unlikely since the
magnitude of the
current is evidently not  sensitive to whether twins are
present.

To the extent that twinned YBCO can be treated as a tetragonal
system, these two
types of experiments appear manifestly incompatible. $\pi$ phase
shifts are
associated with a sign change of the order parameter in the $a,b$-
plane.
The $c$-axis
current, which  reflects an average of the order parameter in the
$a,b$-plane, must
necessarily vanish.  These observations are not altered when bi-layer
effects are
incorporated.  While there has been some attention
\cite{MazGolZai,Pokrovski,Combescot} paid to the role of
chains in addressing the data, it should be noted that recent
experiments on
reduced oxygen YBCO \cite{BrawnOtt} (where the chains are expected to be
highly
interrupted and, therefore, irrelevant to the transport and
superconductivity),
seem to reproduce the same behavior as in the optimally doped
material. In summary, if both $a,b$- and $c$-axis
Josephson measurements prove to be correct, any
resolution of
this issue will probably revolve around a  deeper understanding of
twinning effects.

In a Josephson experiment, the measured Josephson current is the
sum of the currents established between each band of the
cuprate and the superconductor to which it is coupled
\begin{equation}
J = J^{+} + J^{-}
\label{sum-J}
\end{equation}
where each component has the usual Ambegaokar-Baratoff form \cite{AmbBara}
\begin{equation}
J^{\pm} = \Delta_{\pm}^L \Delta^R \frac{\pi}{\beta r_{\pm}}
\sum_n \left[(\omega_n^2 + {\Delta_{\pm}^L}^2)
(\omega_n^2  + {\Delta^R}^2)\right]^{-\frac{1}{2}}
\end{equation}
The resistances, $r_{\pm}$, can be calculated by determining
the single particle tunneling matrix elements
\cite{Wolf}
\begin{equation}
T_{k,k'}^{\pm} = \int\!\! \mbox{dr}\, \langle\phi_{k,\pm}^L|j(r)|
\phi_{k'}^R\rangle
\label{tunneling-matrix}
\end{equation}
where the integration is carried out over the intermediate
region. The wave-functions $\phi_{\pm}^L$ and $\phi^R$ are the
single particle wave-functions from the left and right
hand side of the junction respectively, and $j(r)$ is
the usual current operator. The fact that  the electrons of the
cuprates are  localized on the copper-oxide
planes requires that they be
described by wave packets with some finite
spread in momentum along the $c$- axis direction.   A microscopic
treatment of
the tunneling process is complicated by a number of issues, which we
cannot
address here: the propagation
of electrons out of a region of localization in the $c$-direction
will lead to scattering effects as the tunneling pair enters a more
isotropic
material. These may influence the tunneling
from both bands to a substantial degree.
(It should also be noted that the order
parameter at the surface may be  modified
from its bulk form.\cite{BuchSauls})
For definiteness, we first consider the case of coupling to a
conventional
superconductor such as Pb, and ignore these complications.

If we make the simplifying assumption that
the Josephson coupling is the same for both the symmetric ($+$) and
antisymmetric
($-$) bands,
it follows that $\pi$ phase shifts in a SQUID experiment
will be observed in a substantial region of the phase space,
corresponding to all
bi-layer parameter sets which lie below the solid line of
Figure~\ref{bilayer-fig}.
Moreover, for a
significant fraction of these (particularly in the vicinity of the van
Hove points),
the states in question differ from two in phase  $d$-like states. They are
described by
$(s,d)$ or $(s,-s)$ dominant combinations.
In this way, the measurement of $\pi$ phase
shifts cannot be uniquely associated with $d$ states in a multi-layer
system,
provided orthorhombicity is also present.  This observation makes it
all the more
important to repeat the various SQUID experiments for either
mono-layer or tetragonal materials.

In reality, there is some asymmetry in the Josephson
coupling to the
symmetric and anti-symmetric bands.  This asymmetry derives from
(1) density
of states effects related to  the van Hove points and (2) wave-
function structure:
wave
functions associated with the anti-bonding band
of the bilayer cuprate have
opposite phase on the two layers and therefore
a nodal plane exists between the layers. The first  effect  tends to
enhance
Josephson coupling preferentially to the antibonding and the second
to the
bonding band. The net contribution  cannot be calculated with any
certainty, but
it is reasonable to conclude that the Josephson coupling is
appreciable for both
bands of the bi-layer system.

These remarks can be addressed in somewhat more detail. In most of
the high
$T_c$ cuprates
for which band structure data is available, the Fermi
surfaces are closed about the $(\pi,\pi)$ point. As a
consequence the anti-bonding band lies closer
to the van Hove points and thus has a higher
density of states.  This density of states contribution will then lead to
a relatively
stronger  Josephson coupling  for the anti-bonding or antisymmetric
band.  On
the other hand, the amplitude
of the pair wave function is expected to be greater on the even
symmetry band \cite{fn-ref-to-fig}
and this effect may lead to  a greater
contribution to the Josephson current than for the even band.
Moreover, it has
been claimed \cite{TMRice} on the basis of symmetry arguments,
that tunneling
from
the anti-bonding band  into an
$s$-wave superconductor will lead to an appreciable reduction in the
current
contribution from this band. However, some Josephson coupling is
expected to
remain; the matrix elements $T_{k,k'}^-$ only vanish  under the special
circumstances, when the centers of the two
wavefunctions of Eq.~(\ref{tunneling-matrix})
co-incide throughout the boundary region. In general, this
matrix element will be nonzero, if for no other
reason than because the lattice constants of the two
materials are unlikely to be commensurate.
The situation is depicted in Figure~\ref{tunnel-fig}.

Josephson tunneling from one YBCO crystal to another, as well as
across twin
boundaries,   is  even less amenable to microscopic theory.
In many respects twinned materials
behave like single crystals, with similar transition
temperatures, thermodynamics and Josephson currents in
corner and $c$-axis junctions. It is clear that little is understood at a
detailed
theoretical level  about the nature  of
the twin boundary. What seems to be less ambiguous, however, is
that there is
little if any pinning of the critical current (in low magnetic fields) as
it flows
between twin boundaries.\cite{Welp}  This would suggest that,
whatever the
order parameter symmetry, the phases tend to line up with $+$ and
$-$ lobes adjacent.

In the context of a bi-layer system, this situation is more complex,
since the
phases are associated with multiple bands. In Figure~\ref{twin-fig},
we
schematically plot the  two alternative scenarios for the $(s,-s)$
(orthorhombic) bands (parts (b) and (c)),
as well as the generally expected behavior for the $(d,d)$ case
(a).  Of the two scenarios (b) and (c), only the latter would preserve
the $\pi$
phase shift behavior across a twin boundary. This scenario would
also be
compatible with a low twin boundary pinning of the critical current.

Despite these phenomenological arguments,  case (b) cannot be ruled
out on
microscopic grounds.  In the extreme limit  where the boundary can
be
treated
as a conventional SIS junction,  the locking of the
phase of the order parameter across the junction occurs via
Josephson coupling. If the bilayers on either side of the
junction are properly aligned it might be expected that only bands
of equal
symmetry couple together
since
the matrix element in Eq.~(\ref{tunneling-matrix}) vanishes
otherwise.
Thus
the even bands on either side of the junction would couple
together as would the odd bands.
As
a consequence the positive lobes of a $(\Delta_{+},\Delta_{-
})=(d,d)$
solution would point in the same spatial direction on
both sides of the twin boundary(scenario (a)). On the other hand the
$(s,-s)$
state would have the component of positive
phase in one twin aligned with the component with
negative phase in the other twin (scenario (b)). This would
lead to a substantially reduced Josephson current at a
macroscopic junction due to averaging over the twins. If instead of
an SIS model,
one argued that the order parameter lobes were required to vary
in the most
continuous fashion, one might  conclude that the dominant
component
$\Delta_{\perp}$ should also be continuous. This, too, would lead to
scenario (b) for
the $(s,-s)$ states and (a) for the $(d,d)$ configuration.

Nevertheless, it is also
possible to assume that interlayer pair breaking effects
become considerable in the boundary region and that
$\Delta_{\parallel}$ retains its coherence between twins so that
scenario (c)
obtains.
Moreover, stacking faults, lattice defects and
other complexities  can invalidate any of the above simple models
of twinned crystals. In summary, the nature of the order parameter
variation
across a twin boundary is quite complicated  in one layer systems
and
sufficiently complex in the bi-layer case so that no clear conclusions
can be
drawn at this time.

\subsection{Landau-Ginzburg Free Energy Functional}
\label{LG}

Thus far we have investigated only the linearized gap
equations, which are necessarily restricted to  the vicinity
of the transition temperature.  Additional effects may occur
below $T_c$ associated with the transition to states with
other order parameter symmetries. While there does not
appear to be experimental evidence for additional phase
transitions, there is considerable information contained in
studying the more general situation.  In this section we
derive the appropriate Landau-Ginzburg free energy
functional in terms of both the layer and band indices. The
behavior at quadratic order gives further insight into, and
serves to validate the results discussed in the previous
sections. There have been several discussions in the literature
\cite{KobukiLee,Varma} of phenomenological forms for
the bi-layer free energy.  Here we proceed from a microscopic basis.
Note that while this discussion
refers to a bilayer structure, it is readily extended to the
case of general $N$ following the results of
Section~\ref{Ideal}.

In the case under consideration (when only pairing of electrons on
individual sub-bands
of the Fermi surface is considered), the superconducting
state of the bi-layer is described by a two component order
parameter : $\Delta_+$  and $\Delta_-$.  The free energy, $F_s$,
of the superconducting
state then consists of the sum of terms
\begin{equation}
F_s = F_{s,+} + F_{s,-}
\label{Fs}
\end{equation}
where $F_{s,\pm}$ is the contribution to the free energy of an
individual sub-band. In the Landau-Ginzburg limit
the free energy difference
between the normal and superconducting states is given by
\begin{equation}
F_{s,\pm}-F_{n,\pm}\:=\:\ln\frac{T}{T_c}
\int_{\epsilon_{\pm}=E_F}\!
\frac{\mbox{dS}}{(2\pi)^2v_F}\left|\Delta_{\pm}\right|^2
\:+\: \frac{0.0533}{(k_BT_c)^2}
\int_{\epsilon_{\pm}=E_F}\!
\frac{\mbox{dS}}{(2\pi)^2v_F}\left|\Delta_{\pm}\right|^4
\end{equation}
It should be stressed that, although the two sub-bands contribute
independently to the free
energy, the two order parameters  are coupled via the gap
equation. Both become finite at a common $T_c$.

By use of the gap equations the the Landau-Ginzburg free energy
functional can be recast into a form
which contains a quadratic part
\begin{equation}
\alpha_+|\Delta_+|^2 +
\alpha_-\left|\Delta_-\right|^2 +
\delta\left(\Delta_+\Delta_-^* + c.c. \right)
\label{LG-quad}
\end{equation}
and a  quartic contribution
\begin{equation}
\beta_+|\Delta_+|^4 + \beta_-|\Delta_-|^4
+ \left(\Delta_+\Delta_-^* + c.c. \right)
\left(\mu_+|\Delta_+|^2 +
\mu_-|\Delta_-|^2\right)
\label{LG-quart}
\end{equation}
Here we have assumed that each component of the order
parameter can be separated into a complex magnitude
times a normalized real
function over the Fermi surface:
\begin{equation}
\Delta_{\pm}({\bf q}) =
\Delta_{\pm}\psi_{\pm}({\bf q})
\label{q-sep-LG}
\end{equation}
The magnitude $\Delta_{\pm}$
appears explicitly in the Landau-Ginzburg free energy expansion
while $\psi_{\pm}$ determines the co-efficients
in the expansion.
The co-efficients $\delta$ and
$\mu_{\pm}$ vanish if $\Delta_+$ and $\Delta_-$ belong
to different irreducible representations of the lattice point group.

These coefficients have been discussed by Varma
in the context of
a particular
pairing scenario.\cite{Varma} While our free energy contains
the same class of terms as that presented in Ref.~\onlinecite{Varma},
because we have explicitly removed the wave vector dependences via
Eq.~(\ref{q-sep-LG}) above, it is not straightforward
to determine the conditions for
order parameter sign changes under a $\pi /2$ rotation of the lattice.
Nevertheless, the relative phase of the  two order parameters can be
determined variationally, by minimizing the free energy.
If the nontrivial coupling parameters ($\delta$
and $\mu_{\pm}$) are finite then the phase difference between
$\Delta_+$ and $\Delta_-$ can be at most $0$ or $\pi$.
This last result is consistent with the discussion of
Section~\ref{2l-solutions} and the appropriate sign
depends on whether in-plane or interplane correlations are
dominant.\cite{fn-LGterms}

It is useful to transform the above free energy functional to  the layer
representation using Eq.~(\ref{band-to-layer}).
The functional is expressed in terms of  $\Delta_{\parallel}$
and $\Delta_{\perp}$
and two explicit transition temperatures can then be associated with
the intra- and inter-plane interactions. In this context
the mixing between various symmetries
can be understood in a more direct way.
In this representation the free energy is given by
\begin{equation}
\begin{array}{c}
\alpha_{\parallel}|\Delta_{\parallel}|^2 +
\alpha_{\perp}\left|\Delta_{\perp}
\right|^2 +
\delta'\left(\Delta_{\parallel}\Delta_{\perp}^* + c.c. \right) +
\beta_{\parallel}|\Delta_{\parallel}|^4 +
\beta_{\perp}|\Delta_{\perp}|^4 +
\beta'|\Delta_{\parallel}|^2 \left|\Delta_{\perp}\right|^2 \\
+ \gamma'\left(\Delta_{\parallel}^2 \Delta_{\perp}^{*2} + c.c.
\right) + \left(\Delta_{\parallel}\Delta_{\perp}^* + c.c. \right)
\left(\mu_{\parallel}|\Delta_{\parallel}|^2 +
\mu_{\perp}|\Delta_{\perp}|^2\right)
\end{array}
\label{LG-layer}
\end{equation}

Here the coupling terms $\delta'$ and
$\mu_{\parallel/\perp}$
vanish if $\Delta_{\parallel}$
and $\Delta_{\perp}$ belong to different irreducible representations
of the lattice symmetry group.
Note that one important effect of orthorhombicity
is to require that  the coupling parameters $\delta'$ and
$\mu_{\parallel/\perp}$ change sign upon the interchange
of the $a$ and $b$ axes of the crystal.
These same terms also vanish in the limit as $t_{\perp}$ goes to
zero.

By direct calculation from the microscopic theory
we find that the quartic cross-term $\gamma'$
is positive \cite{fn-show-gamma'}
so that the  phase difference between
$\Delta_{\parallel}$ and $\Delta_{\perp}$ is zero (if $\delta' < 0$)
or $\pi$ (if $\delta' > 0$).\cite{fn-delta=0}
Studies of a related Landau-Ginzburg free energy functional
have been
presented by Kobuki and Lee \cite{KobukiLee} who
discussed the mixing between a $d$-wave
$\Delta_{\parallel}$ and an $s$-wave $\Delta_{\perp}$
in an RVB based theory.
In their approach, mixing between these two components was brought about
by a self-consistently determined orthorhombic strain.
This led to the introduction of additional terms into the free energy
which could
result in a negative $\gamma'$. As a consequence
an $s+\iota d$ state was produced.
In the present theory, in contrast, mixing
between $\Delta_{\perp}$ and $\Delta_{\parallel}$ occurs at quadratic
rather than quartic order, as a consequence of finite
$t_{\perp}$. Furthermore, in the absence of
tetragonal symmetry breaking
only $s+d$ mixing occurs (at quartic order).
Additional arguments against an $s+\iota d$
state were presented by Normand et al.\cite{Normand}

\subsection{Impurity Effects}
\label{dirt}

In this section we discuss the nature of impurity or pair-breaking
effects in bi-
layer systems. At the heart of this issue is the  paradoxical
observation that all
substitutions at the rare earth site, which sits between the bi-layers
(except for
Pr),  leave $T_c$ unaffected.  Rare earth substitutions with or
without local
moments and in a disordered  or ordered form make no difference to
the
superconducting transition temperature.  Previously it has been
argued that the even more
general insensitivity of the cuprates to impurity substitution
is incompatible with
anisotropic or $d$-wave superconductivity.
\cite{Millis,RadLevin-dirt,Sun-prb50}
Here we
investigate the complexity introduced into this problem by  the
presence of a bi-
layer order parameter.  Since it is relatively straightforward to
generalize to the
magnetic case, for definiteness, we concentrate on the case of non-
magnetic
impurities.

A new aspect of the present work is the consideration of an isolated
bi-layer,
rather than a coherent stack of bi-layers. Our starting point is a
necessary first
step in a treatment of incoherent coupling  along the $c$-axis.  As a
result of this
assumption the configuration averaging process (which restores the
underlying
translational symmetry of the lattice) is different from that discussed
in Refs.~\onlinecite{KetEf,KlemmScharn}.
As has been noted  elsewhere,\cite{KetEf}  there are two
types of
impurities which must be considered: intra and inter-layer
substitutions.  For an
isolated bi-layer, inter-layer impurities represent the more
interesting case, since
the relevant equations  decouple when written in the band basis.
Thus a state
such as an isotropic  $(s,-s)$ state will be insensitive to inter-layer
impurities
(as will the in-phase $(s,s)$ state). By contrast, intra-layer impurity
effects involve
processes which couple the two bands. In this way, the $(s,-s)$
state exhibits
intra-layer pair-breaking.\cite{KetEf} Moreover, the same concerns
that were
raised earlier \cite{RadLevin-dirt,Sun-prb50} about the $d$-wave order
parameter apply to
the bi-layer case with either type of impurity. Thus the $(s,-s)$ state
emerges as the
leading (non-trivial)  candidate state for resolving the paradox
concerning rare
earth (non-magnetic) substitutions in the cuprates.

We begin with the standard treatment of  scatterers within
individual planes
using the Born approximation.\cite{Maki}
The impurity Hamiltonian has the form
\begin{equation}
\hat{H}_{\scriptbox{in-plane}}^{\scriptbox{imp}}
= \sum_m \sum_{{\bf q},\sigma } u_I({\bf q})
c_{i_m,\sigma }^\dagger({\bf p}+{\bf q}) c_{i_m,\sigma }({\bf p})
e^{-\iota{\bf q \cdot R}_m}
\end{equation}
where $m$ labels impurities located at ${\bf R}_m$ on layer
$i_m$
and $\sigma$ is a spin index. The impurity self-energy is given by
\begin{equation}
\Sigma_{\omega,r,r'}({\bf p}) = -\frac{1}{2}n_{\scriptbox{imp}}
\sum_{r",{\bf q}} \left|u_I({\bf p}-{\bf q})\right|^2
\bar{\cal{G}}_{\omega,k"}({\bf p})\delta_{r,r'}
\end{equation}
where $\bar{\cal{G}}$ is  the averaged  Green's function
and $n_{\scriptbox{imp}}$  the concentration of
impurities per layer. We assume that the gap in the absence of
impurities is given
by $\Delta_\pm({\bf q})
=\bar{\Delta}_\pm\psi_\pm({\bf q})$ where $\psi_\pm$ are
functions normalized appropriately over the corresponding
sub-band of the Fermi surface and that the impurity renormalized
gap is given by $\tilde{\Delta}_\pm\psi_\pm({\bf q})$.
The self consistent equations then become
\begin{mathletters}
\label{self-cons-I}
\begin{eqnarray}
\tilde{\omega}_l & = & \omega_l +
\frac{1}{4\tau}\sum_{r'=\pm}
\frac{\tilde{\omega}_l}{\sqrt{\tilde{\omega}_l^2
+ \tilde{\Delta}_{r'}^2}} \\
\tilde{\Delta}_\pm & = & \bar{\Delta}_\pm +
\frac{g}{4\tau}\sum_{r'=\pm}\frac{\tilde{\Delta}_{r'}}
{\sqrt{\tilde{\omega}_l^2 + \tilde{\Delta}_{r'}^2}}
\end{eqnarray}
\end{mathletters}
where $\tau$ is the usual scattering time and
\begin{equation}
\frac{g}{2\pi\tau}=\sum_{r=\pm}\sum_{r'=\pm}
\int\!\!\frac{\mbox{dS}_r}{(2\pi)^2v_F^r}
\int\!\!\frac{\mbox{dS}_{r'}}{(2\pi)^2v_F^{r'}}\,
u_I^2({\bf p}-{\bf p}')\,\psi_r({\bf p})\,\psi_{r'}({\bf p}')
\end{equation}
Here $g$ is a combined measure of  the anistropy of the order
parameter\cite{Millis} and
impurity potential.  This  coupling constant varies from $0$ to
$1$.  The latter
is appropriate to the case of  a totally isotropic order
parameter.

It follows from Eqs.~(\ref{self-cons-I}) that intra-layer
impurities enter the coupled self consistent equations via a mixture of the two
band
contributions.
In this way they lead to pairbreaking in all instances, except for the
special case
of two in phase, isotropic $s$-states. Thus the $(s,-s)$ states experience
a reduced
$T_c$ in the presence of these impurities.\cite{fn-Tc-sup-1}

We next consider  inter-plane scatterers. For the sake of generality
these
impurities are assumed to scatter electrons within as
well
as between planes.
The impurity Hamiltonian is given by
\begin{equation}
\hat{H}_{\scriptbox{interplane}}^{\scriptbox{imp}}
= \sum_m \sum_{{\bf q},\sigma} \left[
u_{\parallel}\left(c_{1,\sigma}^{\dag}c_{1,\sigma}
+c_{2,\sigma}^{\dag}c_{2,\sigma}\right)
+u_{\perp}\left(c_{1,\sigma}^{\dag}c_{2,\sigma}
+ c.c.\right)\right]e^{-\iota{\bf q \cdot R}_m}
\label{imp-def}
\end{equation}
It is important to note that this Hamiltonian
is diagonal in the band language and that no average is taken over
sites in the
vertical direction.  Thus the   renormalized self-energy and order
parameter satisfy  equations which are decoupled in the band index
\begin{mathletters}
\label{self-cons}
\begin{eqnarray}
\tilde{\omega}_{l,\pm} = \omega_l + \frac{1}{2\tau_\pm}
\frac{\tilde{\omega}_{l,\pm}}{\sqrt{\tilde{\omega}_{l,\pm}^2
+ \tilde{\Delta}_\pm^2}} \\
\tilde{\Delta}_\pm = \bar{\Delta}_\pm +
\frac{g_\pm}{2\tau_\pm}\frac{\tilde{\Delta}_\pm}
{\sqrt{\tilde{\omega}_{l,\pm}^2 + \tilde{\Delta}_\pm^2}}
\end{eqnarray}
\end{mathletters}
Here care must be taken to preserve the band labels on $\tau$ and
$g$.
This decoupling of the two bands leads to the conclusion, stated
earlier, that
isotropic order parameter sets (whether in or out of phase)
experience no
pairbreaking from inter-layer impurities.\cite{fn-Tc-sup-2}

In summary, we see that one way of avoiding the strong
pair-breaking generally associated with substitutions
at the rare earth site between the bi-layers is to consider
states with the symmetry $(s,-s)$.
For this reason, along with the $\pi$ junction
behavior discussed in the previous section, this state
should be considered as a potentially interesting candidate
for the YBCO system.

\section{Formulation and Analysis of the $N$-Layer Problem}
\label{Analysis}

\subsection{Gap Equation in the Band Representation}
\label{Calc-Gap}

In this section, we treat the general $N$-layer problem. The system
under consideration corresponds to  a stack of decoupled $N$-layer
structural units, each layer of which consists of a two dimensional
copper oxide plane.  In the limit of infinite $N$, we recover essentially
the usual
Bloch wave description of a collection  of copper oxide mono-layers,
aligned along the $c$-axis.\cite{KlemmLiu,KetEf}
Just as in the bi-layer case, we
find that there are two competing
states. Depending on the relative size of the intra- and inter-plane
coupling constants; one  of the two is  stable, while the other is
metastable.  These two limits correspond to inter- and intra-layer
dominated states. In the band language the latter are in phase and
the former out of phase.  Because the system contains $N$ such bands,
the inter-layer gap parameters should be viewed as sinusoidally
modulated with varying band index, as will be illustrated in more
detail below.

Just as in the bi-layer case, the charge carriers within individual
planes and on
adjacent planes within a unit cell are assumed to interact via a
non-
retarded
pair potential. Hopping of
quasiparticles
between adjacent planes within a unit cell is
determined by the hopping matrix element $t_{\perp}$. No hopping
is allowed between unit cells, as a consequence of our assumption
that the $c$-axis coupling is incoherent.
As before,  only singlet intraband pair states are considered.
Many of the detailed derivations in this section may be found
in Appendix~\ref{derive-app}.

The non-interacting Hamiltonian has the form
\begin{equation}
\hat{H}_N = \sum_{\sigma }\left[\sum_{i=1}^N
\xi({\bf q})c^{\dag}_{i\sigma }c_{i\sigma }^{} -
\sum_{i=1}^{N-1}\left(t_{\perp}c^{\dag}_{i\sigma }c_{i+1,\sigma }^{}
+ c.c.\right)\right]
\label{H_0}
\end{equation}
where $i$ is a layer and $\sigma $ a spin index.
The superconducting order parameter is defined by
\begin{equation}
\Delta_{ij}({\bf q}) =
\sum_{{\bf q}'}V_{ij}({\bf q}, {\bf q}')F_{ij}({\bf q}')
\label{Delta_ij-def}
\end{equation}
where $V_{ij}$ is the interaction between electrons on layer
$i$
and layer $j$ and the anomalous Green's function $F_{ij}({\bf
q}) =
\langle c_{i\uparrow}({\bf q})c_{j\downarrow}(-{\bf
q})\rangle$
is antisymmetrized with respect to spin indices.

The
anomalous component
of the superconducting Green's function is obtained by
performing the
usual matrix inversion
\begin{equation}
{\cal G}^{-1} = \left( \begin{array}{cc}
\iota\omega_l - H_N   &  -\Delta  \\
-\Delta^{*}  &  \iota\omega_l + H_N
\end{array} \right)
\label{m-eq1}
\end{equation}
The entries in this matrix problem are $2N \times 2N$
matrices with
two spin
degrees of freedom and $N$ layer indices.

We diagonalize Eq.~(\ref{m-eq1}) in the spin degrees of freedom by
multiplying
the entire equation from the right by $\sigma_2$.
The equations are then transformed to the band
picture by diagonalizing the normal state Hamiltonian $H_N$.
The resulting energy dispersion in band $r$ of the
Hamiltonian
is given by
\begin{equation}
\epsilon_r({\bf q}) = \xi({\bf q}) +
2t_{\perp}\cos\!\left(\frac{r\pi}{N+1}\right)
\ \ \ \ \  \mbox{$r = 1,\ldots, N$}
\label{e-disp}
\end{equation}
We will consider only pairing of electrons within individual
bands so that the order parameter is diagonal in the
band representation with components $\Delta_r$.
With a fully diagonalized Green's function the gap equation
for
the $N$-layer system can now be readily obtained. Upon
performing
the sum over Matsubara frequencies, we find
\begin{equation}
\Delta_r({\bf q}) = -\sum_{{\bf q}'}
\sum_{r'=1}^{N} V^{r,r'}({\bf q}, {\bf q}')
\frac{\Delta_{r'}({\bf q}')}{2E_{r'}({\bf q}')}
\tanh\!\left[\frac{1}{2}\beta E_{r'}({\bf q}')\right] \\
\label{Gap-eq}
\end{equation}
The energy dispersion of the elementary excitations
is given for each band $r$ by the usual relation
$E_r=\sqrt{\epsilon_r^2+\left|\Delta_r\right|^2}$.
Equation (\ref{Gap-eq}) is the central equation of this section. In
Appendix \ref{3D-app} we generalize this result
further by extending it to an infinite stack of layers, corresponding to
a fully three dimensional lattice.

\subsection{Solutions of the Gap Equation for Small $t_{\perp}$}
\label{Ideal}

In this sub-section we establish the nature of the two competing
states which are the stable and metastable solutions to the gap
equations derived from Eq.~(\ref{Gap-eq}). These are most
readily introduced by considering first the limit of small $t_{\perp}$.
For arbitrarily small  $t_{\perp}$, the two solutions become
independent and appear with different onset or transition
temperatures.  These two states  are respectively associated with
pure intra- and pure interlayer pairing.  Moreover, in this limit analytical
results can be obtained, while the more general case of non-zero
$t_{\perp}$ is treated numerically.

After some algebra, which is outlined in Appendix \ref{uncouple-app},
it follows that the solution to the gap equation is given by
\begin{equation}
\Delta_r({\bf q}) = \Delta_{\parallel,0} \ \psi_{\parallel}({\bf
q})
- 2\cos\!\left(\frac{r\pi}{N+1}\right)
\Delta_{\perp,0} \ \psi_{\perp}({\bf q})
\label{ideal-delta}
\end{equation}
where the two Fermi surface functions $\psi_{\parallel}({\bf q})$
and $\psi_{\perp}({\bf q})$ satisfy
\begin{mathletters}
\label{1l-ideal}
\begin{eqnarray}
\Omega_{\parallel}\,\psi_{\parallel}({\bf q}) & = &
-\int\!\!\frac{\mbox{dS}}{(2\pi)^2v_F}\,
V_{\parallel}({\bf q}, {\bf q}') \,\psi_{\parallel}({\bf q}') \\
\Omega_{\perp}\,\psi_{\perp}({\bf q}) & = &
-\int\!\!\frac{\mbox{dS}}{(2\pi)^2v_F}\,
V_{\perp}({\bf q}, {\bf q}') \,\psi_{\perp}({\bf q}')
\end{eqnarray}
\end{mathletters}
Here the integrations are over the degenerate bands of the
Fermi surface and the related c-numbers, $\Omega_{\parallel}$
and
$\Omega_{\perp}$, are related to the respective transition
temperatures  $T_{\parallel/\perp,0}$ determined by the
equation
\begin{equation}
\ln\!\left(\frac{1.14\,\omega_c}{k_B\, T_{\parallel/\perp,0}}\right) =
\frac{1}{\Omega_{\parallel/\perp}}
\end{equation}
Finally, the two competing states are associated with taking either
one of the  two parameters in $\Delta_{\parallel,0}$
or $\Delta_{\perp,0}$ in Eq.~(\ref{ideal-delta})
to be zero.

The above results can be generalized to the case of finite $t_{\perp}$,
using numerical techniques. When $t_{\perp}$ is finite, the two
parameters $\Delta_{\parallel,0}$ and  $\Delta_{\perp,0}$ can be
simultaneously  non-zero in which case both
$\psi_{\parallel}$ and $\psi_{\perp}$ will belong to the same
irreducible representation of the lattice point group.

To illustrate these results, we plot the amplitudes on the various
bands of the competing
inter- and intralayer
pair states for $N=4$ and $N=7$  in
Figure \ref{nlayer-fig}.
As in the bi-layer case,  if the
dominant interaction is attractive
then the order parameter will be nodeless, while
a repulsive interaction will yield a nodal solution
with exact form determined by the details of
the interaction and the band structure
(as discussed in Section~\ref{1l-solutions}).
In panels (a) and (b) of
Figure \ref{nlayer-fig} the magnitude of the order parameter is
plotted as a function of band index $r$   for
the intraband (a) and interband (b) states for a four layer system.
The dotted line indicates the analytical solution for
$t_{\perp} = 0 $ and the histogram bars illustrate the numerical
results for moderate $t_{\perp}$ (comparable in magniude to the separation
of the Fermi level from the van Hove points).
In this way some deviation from
the analytically obtained curve is seen as the solid bars differ
slightly from the dotted line.  It is clear that the two competing
solutions represent a natural generalization of the bi-layer results to
an $N$-layer system.  Similar results are plotted for the seven layer
system in panels (c) and (d).  It follows  from the figures and the
above discussion that in these higher $N$ systems, even more
complex behavior can be obtained, with a range of signs and
magnitudes of the order parameters associated with the different
bands.

To make this complexity even more explicit, we have considered the
case of $N=3$ for the case of a dominant attractive interlayer interaction
and a weak in-plane repulsion both peaked at $(\pi,\pi)$.
In this case the hopping is slightly larger than in the previous
figure.
Figure \ref{3l-fig}
shows our solution to the gap equation and the normal state
bandstructure (inset)  for this 3 band model of orthorhombic YBCO with
the two lower energy bands closed about $X$
and the highest energy band open. This
last band can be viewed as simulating the  chain band in YBCO,
\cite{MazGolZai}  since conduction in this band is only
possible along one principle axis. Solution of the 3 layer gap equation
clearly shows the mixing between the two components
of the solution. The middle (plane-like) band
has a pure $d$-wave
solution,  since interlayer pairing contributes very little
to the gap on this band, while the other two bands have $s$-wave
symmetry.   The physics of this 3 band model is equally complex.
Because of the dominance of an  $s$-wave order parameter
component, one expects that the magnitude of $T_c$ is
only mildly affected by impurity scatterers.
On the other hand, a nodal solution
on one of the bands  will yield power law dependences in
thermodynamic functions at low temperatures.  While the above model should not
be viewed as a detailed representation of YBCO, it serves to illustrate the
rich array of phenomena which are associated with multiband systems.

\subsection{Basis Functions and Van Hove Effects}
\label{van-Hove}

Band structure effects have played an important role in our analysis,
particularly when the Fermi energy lies in the vicinity of the van
Hove singularities. We have seen that these singularities distort the
shape of the order parameter.
They also play a key role in
determining the relative stability of various solutions to the gap
equations.  In this sub-section we show that, of all the different gap
symmetries, two are able to take maximal  advantage of the van
Hove points: these are the nodeless $s$ if the interaction is
attractive and the "$d_{x^2-y^2}$" states\cite{TsueiNewns-up95}
in the case of a repulsive
interaction.  It should be noted that there is a considerable literature
on the effect of the van Hove singularities on raising $T_c$.
\cite{NewnsTsuei-cmp15}
In this paper we focus on the interplay of the van Hove singularity
and order parameter symmetry. Earlier work
\cite{RadNorm,RadLevin-vanHove} has shown that
in the more general strong coupling picture, states far from
the Fermi energy may wash out, to some degree, the effectiveness of the
van Hove singularity in raising $T_c$. While here we use a weak coupling
approach to address the order parameter symmetry
the same qualitative behaviour can be expected to
follow in more general strong coupling calculations.

To quantify the van Hove effects we study the linearized form of Equation
(\ref{Gap-eq})
\begin{equation}
\Omega \, \Delta_r({\bf q}) =
- \sum_{r'=1}^N \int_{\epsilon_{r'}=E_F}\!
\frac{\mbox{dS}_{r'}}{(2\pi)^2v_F^{r'}({\bf q}')}\,
V^{r,r'}\!({\bf q}, {\bf q}')\,\Delta_{r'}({\bf q}')
\label{gap-fs}
\end{equation}
where the integrations are performed over segments of the
Fermi
surface corresponding to the different bands
and the eigenvalue $\Omega$ defines the  BCS
transition
temperature $T_c$
\begin{equation}
\Omega^{-1} = \ln\! \left(\frac{ 1.14 \, \omega_c}{k_B\,T_c}\right)
\label{Tc-def}
\end{equation}
The Fermi velocity on band $r$ is  $v_F^r$, and
$E_F$ is the Fermi energy.

We define a complete set of orthonormal basis functions over the
Fermi surface,\cite{Allen} which  are non-zero over
only
a single band $r$ and assume further that these
$\psi_i^{r(\Gamma)}({\bf q})$ belong to an irreducible
representation  $\Gamma$ of the lattice point
group. These basis functions
satisfy
\begin{equation}
\sum_{p=1}^N\int_{\epsilon_p=E_F}\!\frac{\mbox{dS}_p}{(2\pi)
^2v_F^p({\bf q})}\,
\psi_i^{r(\Gamma)}\!({\bf q})\,\psi_j^{r'(\Gamma')}\!({\bf q}) =
\delta_{i,j} \, \delta_{r,r'} \, \delta_{\Gamma,\Gamma'}
\label{def-norm}
\end{equation}
The pair wave function and interaction potential \cite{Aoi}  are
expanded in terms of these
Fermi surface harmonics as
\begin{equation}
\Delta_r = \sum_{\Gamma,i} \Delta_i^{r(\Gamma)}
\psi_i^{r(\Gamma)}
\end{equation}
and
\begin{equation}
V^{r,r'}\!({\bf q}, {\bf q}') =
\sum_{\Gamma,i,j}V_{i,j}^{r,r'(\Gamma)}
\psi_i^{r(\Gamma)}\!({\bf q})\,\psi_j^{r'(\Gamma)}\!({\bf q}')
\label{V-irrep}
\end{equation}
The gap equation is thus reduced to the
simple set of eigenvalue problems
\begin{equation}
\Omega\,\Delta_i^{r(\Gamma)} \ = \
-\sum_{j,r'} V_{i,j}^{r,r'(\Gamma)}\Delta_j^{r'(\Gamma)}
\label{eigen}
\end{equation}

By working in the space of functions defined by Eq.(\ref{def-norm})
it is clear that the basis functions $\psi_i^{r(\Gamma)}$
are weighted by the inverse of $\sqrt{v_F^r}$ and so regions along
the directions of the van Hove points give a correspondingly
greater contribution in the gap equation. If the interaction
$V^{r,r'}({\bf q,q'})$ coupling two points on the Fermi surface
is repulsive (attractive) then states with opposite (same) phase
at these two points will be favoured.
Thus we can conclude that the
$d_{x^2-y^2}$ basis function will benefit most from the
van Hove points for repulsive interactions while the nodeless
$s$ function will be most enhanced by an attractive interaction.

The effect of the van Hove singularities in
the single particle density of states on
the superconducting order parameter
can be quantified through the pairing density of states (PDOS)
\cite{FehrenNorm}
\begin{equation}
P_{\Delta}(E_F)
= \displayfrac{\int\!\!\frac{\mbox{dS}}{(2\pi)^2v_F}\,
|\Delta({\bf q})|^2}
{\int\!\mbox{dS} |\Delta({\bf q})|^2}
\label{PDOS}
\end{equation}
The integrals are taken over all bands of the Fermi surface.  To
illustrate this function and its relation to the van Hove singularities,
in Figure~(\ref{PDOS-fig}a)
we plot  $P_{\Delta}$ as a function of the
Fermi energy for various solutions
to the gap equation. Here we focus on the one layer case for clarity.
It can be seen that  a peak appears at the van
Hove point for all irreducible representations. It
is, however,
more appropriate to normalize by the single particle
density of states $P_{\Delta=1}$. This gives a more accurate
indication of the degree to which the pair
wave function is stabilized by the density of states.
The result for the four different irreducible representations
of the $D_{4h}$ lattice of LSCO is plotted against Fermi
energy in Figure~(\ref{PDOS-fig}b). Here  we select that ${\bf Q}$
which results
in a maximal $T_c$ for each representation.  In this way, we see
that the $B_{1g}$ solution, ie. the "$d_{x^2-y^2}$"
state, and the nodeless $A_{1g}$ "$s$"-wave solution indeed
benefit much more from the van Hove
singularity than do all other states.

In summary,  there are two states which take maximal advantage of
the van Hove points. These are the nodeless $s$-wave state
which occurs only for attractive interactions
and the $d_{x^2-y^2}$ state, appropriate to the case of repulsive
interactions.  It should be stressed that in general the order
parameter will not have these simple functional forms corresponding
to a single basis function. This is is all the more striking as
the Fermi energy approaches the van Hove singularity where
admixtures
of higher order basis functions are most evident. While in the
presence of multiple bands,  the results plotted in
Figure~\ref{PDOS-fig} become more complicated,
the essential features still  remain.

\section{Conclusion}
\label{conclusion}

The most important issues in the field of high $T_c$
superconductivity involve determinations of the order parameter
symmetry and the superconducting pairing mechanism. While there are,
clearly, no definitive answers to be had at this time, this paper has been
directed towards addressing  these two issues.  We have emphacized the role of
multi-layer effects in the cuprates, in large part  because the most well
characterized material,  YBCO,  has two copper oxide planes. This
complexity leads to complications in inferring the order parameter
symmetry from various experimental tests.  It also suggests that
there are different (inter- and intra-plane) channels which should be
considered in any microscopic theory of the pairing.

In reference to the order parameter
symmetry, we have found that a multi-layer system should be characterized
by distinct gaps appropriate to each of the multiple bands. A bi-layer
material such as YBCO has two gaps, a tri-layer, three, etc. In the
presence of even a very small amount of orthorhombicity, one of the gaps
can be predominantly  of $d$, while another of $s$ symmetry. (Throughout
this paper we refer to $d$ and $s$ states as those which are odd or even,
respectively, under a $\pi / 2$ rotation of the wave-vector). One may be
node-less while the other has nodes.  The multiple gaps can be in or out of
phase. In this way the observation of power law behavior in
thermodynamics may reflect on only one of the order parameters in
question.  The observation of $\pi$ phase shifts in Josephson corner
junction experiments on YBCO  must be viewed more widely in this multi-
band context. Indeed,  we have found
that this behavior can be associated with
two out of phase $s$-states in the presence of (weak) orthorhombicity.  This
orthorhombicity leads to a strong asymmetry of the $s$ and $-s$ states,
so that one gap function is elongated along the $a$ and the
other along the $b$ axes of the
crystal.  The net Josephson current behaves rather similarly to a $d_{x^2 -
y ^2 }$ state, although the thermodynamical behavior need not exhibit the
power laws of this  state.  Finally, the behavior of multi-layer systems in
the presence of impurities is similarly complex.  Intra- and inter-layer
impurities suppress $T_c$ in a different fashion.  The $s,-s$ state is of
interest because it obeys an Anderson theorem with respect to inter-layer
substitutions.  This may help explain why substitutions at the rare earth
site in YBCO make  little or no difference to the magnitude of $T_c$.

In
the process of investigating very generic model interactions for the
superconductivity, we  have inferred information about microscopic
constraints on the pairing mechanism.  $d_{x^2 - y ^2}$-like states are
found to be general solutions to the gap equation for repulsive
interactions, in large part because they possess  the fewest number of
nodes and thereby the highest transition temperatures.  In this way, they
should not be specifically associated with a spin fluctuation driven
pairing mechanism. Moreover, van Hove effects act to stabilize  some
order parameters over others.  Of these the $d_{x^2 - y ^2}$-like symmetry is,
again,  the most notable.  Orthorhombicity further enhances this
stabilization.  Thus for a variety of reasons, this state emerges as a
natural solution to the gap equation(s) in the presence of repulsive
interactions.

While we have emphacized the bi-layer ($N=2$) case, we
also presented  general multi-layer calculations which view the $c$-axis as
consisting of decoupled structural units, each of which contains  $N$ copper
oxide layers. By contrast, within the unit cell the intra- and
inter-plane hopping is appreciable
and plays an important role in giving rise to $N$ distinct bands. These
$N>2$ calculations may be particularly relevant in the context
of the Hg  and Bi based cuprates. In treating
the superconductivity,  we have included intra- and inter-plane pairing
interactions  in parallel with the above intra- and inter-plane hopping. We
demonstrated that, regardless of the number of layers $N$ in the unit cell,
there are always two competing states: one of which is intra-plane
dominated, so that the resulting $N$ band- gaps are in phase, and one of
which is inter-plane dominated, so that the $N$ gaps are
sinusoidally modulated. Small changes in the parameterizations can lead
to a transition from one of these states to another.  Thus it may be
inferred that the order parameter symmetry is potentially variable from
one cuprate to another and from one stoichiometry to another.

While the inclusion of these multilayer effects has been seen to
introduce considerable complexity into the classification of the
order parameter symmetry, this complexity is inescapable.  As long as
the layers communicate via one or two body processes (i.e., via hopping or
pairing interactions), superconductivity in the high $T_c$ cuprates
must include these multilayer effects.

\emph{Note Added:} After this manuscript was completed we learned of recent
experiments from C. Tsuei and co-workers in which $\pi$ phase
shifts have been reported for a one layer, tetragonal Tl compound.

\acknowledgments

We thank M. Norman and R. Klemm for useful
conversations. This work was supported by the National Science Foundation
(DMR 9120000) through the Science and Technology Centre for
Superconductivity.

\appendix

\section{Calculation of Gap Equation for $N$-Layers}
\label{show-N-layer}

\subsection{Derivation}
\label{derive-app}

To begin we need to calculate the superconducting
Green's function defined by the $2N \times 2N$
matrix
\begin{equation}
{\cal G}^{-1} = \left( \begin{array}{cc}
\iota\omega_l - H_N   &  -\Delta  \\
-\Delta^{*}  &  \iota\omega_l + H_N
				 \end{array} \right)
\label{G-def-app}
\end{equation}
The components $H_N$ and $\Delta$ are defined in Section
\ref{Calc-Gap}. First we diagonalize the normal state
Hamiltonian. For
this
purpose we define the set of characteristic polynomials
\begin{equation}
D_N(\xi - \epsilon) = \det(H_N - \epsilon\,\openone_N)
\label{char-p}
\end{equation}
and observe that they satisfy the recursion relation
\begin{equation}
\begin{array}{rcl}
D_i(x) & = & xD_{i-1}(x) - t_{\perp}^2D_{i-2}(x) \\
D_0(x) & = & 0   \\
D_1(x) & = & x
\end{array}
\label{rec_rel-app}
\end{equation}
By using the known properties of the Chebyshev
polynomials
it is straightforward to show that these characteristic
polynomials have the form
\begin{equation}
\begin{array}{rcl}
D_N(x) & = & \left(-
t_{\perp}\right)^{N}\displayfrac{\sin[(N+1)q]}{\sin(q)}
\\
x & = & -2t_{\perp}\cos(q)
\end{array}
\label{solv-D-app}
\end{equation}
{}From the zeroes of these polynomials
we can determine the normal state energies
\begin{equation}
\epsilon_r({\bf q}) = \xi({\bf q}) +
2t_{\perp}\cos\!\left(\frac{r\pi}{N+1}\right)
\ \ \ \ \ \ \  \mbox{$r = 1,\ldots,N$}
\label{e-disp-app}
\end{equation}
An orthonormal set of eigenstates of $H_N$ can be found
similarly. The $i^{th}$ component of the state associated with
band $r$ is given by
\begin{equation}
\phi_i^r = \sqrt{\frac{2}{N+1}}(-
1)^{i+r}\sin\!\left(\frac{ir\pi}{N+1}\right)
\label{state-comp-app}
\end{equation}
and consequently the components of the unitary matrix
which
diagonalizes (\ref{G-def-app}) are given by
\begin{equation}
U_{ij} = \sqrt{\frac{2}{N+1}}(-1)^{i+j}\sin\!\left(\frac{ij\pi}{N+1}\right)
\label{U_ij-app}
\end{equation}
The particle creation operators in the band and layer
language
are thus related by the equation
\begin{equation}
a_i^{\dag} = \sum_{j=1}^{N}U_{ij}c_j^{\dag}
\label{a=Uc-app}
\end{equation}
Since we predominantly focus on intraband pairing
the order parameter associated with each band $r$ is
thus defined through the anomalous Green's function
components
\begin{eqnarray}
F_{r,r'}({\bf q}) & = & \frac{1}{\sqrt{2}}
\langle a_{r\uparrow}({\bf q}) a_{r'\downarrow}(-{\bf q})
-a_{r\downarrow}({\bf q}) a_{r'\uparrow}(-{\bf q})\rangle
\nonumber \\
& = & \frac{2}{N+1}\sum_{i=1}^N\sum_{j=1}^N
(-1)^{i+j}\sin\!\left(\frac{ir\pi}{N+1}\right)
\sin\!\left(\frac{jr'\pi}{N+1}\right)F_{i,j}({\bf q})\, \delta_{r,r'}
\label{F-trans-app}
\end{eqnarray}
We thus define the $N$ diagonal components of the
order parameter in the band language $\Delta_r$ by the
equation
\begin{equation}
\Delta_r = \frac{2}{N+1}\sum_{i=1}^N\sum_{j=1}^N
(-1)^{i+j}\sin\!\left(\frac{ir\pi}{N+1}\right)
\sin\!\left(\frac{jr\pi}{N+1}\right)\Delta_{i,j}
\label{Delta_r-def-app}
\end{equation}
With the pairing restricted to intralayer pairing and pairing
between nearest neighbour planes we
can use the definition of $\Delta_{i,j}$ to write
the order parameter on a given band $r$ in terms
of the anomalous part of the Green's function
\begin{eqnarray}
\Delta_r({\bf q}) & = & \frac{1}{N+1}\sum_{{\bf q}'}
\left\{\sum_{i=1}^NV_{\parallel}({\bf q}, {\bf q}')
\sin\!\left(\frac{ir\pi}{N+1}\right)
\sin\!\left(\frac{ir'\pi}{N+1}\right)F_{ii}({\bf q}') \right.
\nonumber \\
& & - \left. 2 \sum_{i=1}^{N-1}
V_{\perp}({\bf q}, {\bf q}')\sin\!\left(\frac{ir\pi}{N+1}\right)
\sin\!\left(\frac{(i+1)r\pi}{N+1}\right)
\left[ F_{i,i+1}({\bf q}') + F_{i+1,i}({\bf q}')\right] \right\}
\label{Delta_r-F-app}
\end{eqnarray}
With a fully diagonalized Green's function the gap equation
for
the $N$-layer system can now be readily obtained. Upon
performing
the sum over Matsubara frequencies in the usual way we
get
\begin{equation}
\Delta_r({\bf q}) = -\sum_{{\bf q}'}
\sum_{r'=1}^{N} V^{r,r'}({\bf q}, {\bf q}')
\frac{\Delta_{r'}({\bf q}')}{2E_{r'}({\bf q}')}
\tanh\!\left[\frac{1}{2}\beta E_{r'}({\bf q}')\right]
\label{Gap-eq-app}
\end{equation}
where the interaction in the band representation is defined
as
\begin{eqnarray}
& V^{r,r'}({\bf q}, {\bf q}') = \frac{1}{N+1}\left[
\left(1+\frac{1}{2}\theta_{r,r'}^{+(N)}\right)
V_{\parallel}({\bf q}, {\bf q}') +
\left(2\cos\!\left(\frac{r\pi}{N+1}\right)
\cos\!\left(\frac{r'\pi}{N+1}\right) + \theta_{r,r'}^{-(N)}\right)
V_{\perp}({\bf q}, {\bf q}')\right]
\label{V^rr'-def-app}  \\
& \theta_{r,r'}^{\pm(N)} = \delta_{r,r'} \pm \delta_{r+r',N+1} \nonumber
\end{eqnarray}
where the quasi-particle dispersions in the
superconducting
state are defined by the usual relation
$E_r=\sqrt{\epsilon_r^2+\left|\Delta_r\right|^2}$.

\subsection{The Small $t_{\perp}$ Limit}
\label{uncouple-app}

For infinitesimal $t_{\perp}$
it is interesting to transform
Eq.~(\ref{Gap-eq-app}) to the layer representation
using Eq.~(\ref{Delta_r-def-app}). Then to zeroth order in $t_{\perp}$
the
$N$ intralayer problems uncouple from the $N-1$
interlayer
problems
yielding the following two sets of gap equations
\begin{mathletters}
\label{ideal-app}
\begin{eqnarray}
\Delta_{i,i}({\bf q}) & = & \frac{-1}{N+1}\sum_{{\bf q}'}
V_{\parallel}({\bf q}, {\bf q}')
\frac{\tanh\!\left[\frac{1}{2}\beta E({\bf q}')\right]}{2E({\bf q}')}
\sum_{j=1}^{N} \left( 1+\frac{1}{2}\theta_{i,j}^{+(N)}\right)
\Delta_{j,j}({\bf q}')
\label{ideal-para-app}      \\
\Delta_{i,i+1}({\bf q}) & = & \frac{-1}{N+1}\sum_{{\bf q}'}
V_{\perp}({\bf q}, {\bf q}')
\frac{\tanh\!\left[\frac{1}{2}\beta E({\bf q}')\right]}{2E({\bf q}')}
\sum_{j=1}^{N-1} \left(1+\theta_{i,j}^{+(N-
1)}\right)\Delta_{j,j+1}({\bf q}')
\label{ideal-perp-app}
\end{eqnarray}
\end{mathletters}
Linearizing with respect to $\Delta$ and
solving for the eigenvalues and eigenfunctions of these
two equations gives the $T_c$'s of the various
pairing states available to the system and only one of these
solutions will correspond to the maximal value of $T_c$.

Let us first define two functions on the fermi-surface,
$\psi_{\parallel}({\bf q})$ and $\psi_{\perp}({\bf q})$,
and two numbers, $\Omega_{\parallel}$ and $\Omega_{\perp}$,
which satisfy the two equations
\begin{mathletters}
\label{1l-app}
\begin{eqnarray}
\Omega_{\parallel}\,\psi_{\parallel}({\bf q}) & = &
-\int\!\!\frac{\mbox{dS}}{(2\pi)^2v_F}\,
V_{\parallel}({\bf q}, {\bf q}')\, \psi_{\parallel}({\bf q}')
\label{1l-para-app}    \\
\Omega_{\perp}\,\psi_{\perp}({\bf q}) & = &
-\int\!\!\frac{\mbox{dS}}{(2\pi)^2v_F}\,
V_{\perp}({\bf q}, {\bf q}')\, \psi_{\perp}({\bf q}')
\label{1l-perp-app}
\end{eqnarray}
\end{mathletters}
The integrations are over the degenerate bands of the Fermi surface.
The pair amplitudes can be written in the form
\begin{equation}
\begin{array}{rcl}
\Delta_{i,i}({\bf q}) & = & \Delta_{\parallel}^i\, \psi_{\parallel}({\bf q})
\smallskip  \\
\Delta_{i,i+1}({\bf q}) & = & \Delta_{\perp}^i\, \psi_{\perp}({\bf q})
\end{array}
\end{equation}
The gap equations (\ref{ideal-app})
are transformed in the
usual manner by separating the sum over quasi-momenta
to separate integrals over energy and the Fermi surface.
The energy integral can then be performed to obtain a BCS-like
transition temperature
and using Eqs.~(\ref{1l-app})
we derive the linear matrix equations
\begin{mathletters}
\label{vec-app}
\begin{eqnarray}
\Delta_{\parallel}^i  & = & \ln\!\left(\frac{\gamma\, \omega_c}{k_BT_c}\right)
\frac{\Omega_{\parallel}}{N+1}\sum_{j=1}^{N}
\left(1+\frac{1}{2}\theta_{i,j}^{+(N)} \right) \Delta_{\parallel}^j
\label{vec-para-app} \\
\Delta_{\perp}^i & = & \ln\!\left(\frac{\gamma\, \omega_c}{k_BT_c}\right)
\frac{\Omega_{\perp}}{N+1}\sum_{j=1}^{N-1}
\left(1+\theta_{i,j}^{+(N-1)}\right)
\Delta_{\perp}^j
\label{vec-perp-app}
\end{eqnarray}
\end{mathletters}
where $\gamma \approx 1.14$ and $\omega_c$ is the usual cutoff
energy. We can now characterize all the available pairing states with
functional form given by $\psi_{\parallel}$ and $\psi_{\perp}$
along with their BCS transition temperatures.

Eqs.~(\ref{vec-app}) each
have one isotropic eigenvector given by
$\Delta_{\parallel/\perp}^i=\Delta_{\parallel/\perp,0}$ with
$T_{c,0}$ determined by the equation
\begin{equation}
\ln\!\left(\frac{\gamma\, \omega_c}{k_BT_{c,0}}\right) =
\frac{1}{\Omega_{\parallel/\perp}}
\end{equation}
which give the
most stable candidate pairing states for the two mechanisms.
The remaining eigenvectors correspond either to metastable
states which are symmetric in the layer index such that
$\sum_i\Delta_{\parallel/\perp}^i = 0$ with $T_c$ given by
\begin{equation}
\ln\!\left(\frac{T_c}{T_{c,0}}\right) = \left\{
\begin{array}{c}
-\displayfrac{N}{\Omega_{\parallel}} \\ -\displayfrac{N-1}{2\Omega_{\perp}}
\end{array}
\right.
\end{equation}
or non-pairing states ($T_c=0$) which are odd in the layer index.

Transforming to the band
picture using Eq.~(\ref{Delta_r-def-app}) we find that
the intralayer pairing states are even under the
transformation $r \rightarrow N+1-r$ whereas the
interlayer pairing states are odd under this
transformation. The two most stable candidate states
thus give a superconducting order parameter of the
form
\begin{equation}
\Delta_r({\bf q}) = \Delta_{\parallel,0} \ \psi_{\parallel}({\bf q})
- 2\cos\!\left(\frac{r\pi}{N+1}\right)
\Delta_{\perp,0} \ \psi_{\perp}({\bf q})
\end{equation}
The two parameters $\Delta_{\parallel,0}$ and $\Delta_{\perp,0}$
are both nonzero only when $t_{\perp}$ is finite and when
$\psi_{\parallel}$ and $\psi_{\perp}$ belong to the same
irreducible representation.
Since the two different pairing mechanisms give solutions
of different $r$ dependence, the dominant type of
pairing can be easily determined even when $t_{\perp}$ is finite.
Mixing of solutions at finite $t_{\perp}$ and the effect
on the transition temperature is discussed in Appendix~\ref{IB-pairing}
in the bi-layer context.

\subsection{Formulation In Terms of Bloch Waves}
\label{3D-app}

For completeness we conclude by relating the above
formulation
to the usual treatment of layered materials.
It would be natural to define a $z$-component of the
quasi-momentum
vector to be
\begin{equation}
q_z=\frac{r\pi}{N+1}
\end{equation}
Note, however, that there are $N$ linearly independent,
non-degenerate eigenstates for \mbox{$0<q_z<\pi$}
and so taking \mbox{$q_z \rightarrow -q_z$}
gives no new states. Thus interpreting $q_z$ as a
momentum is rather unnatural.

The usual procedure in the case of large $N$,\cite{KlemmLiu,KetEf}
however, is to
assume periodic boundary conditions. This means that we
have
hopping between layers $1$ and $N$ and that the system is
translationally invariant along the $c$-axis.
The eigenstates of this new Hamiltonian are
\begin{equation}
\phi_j^r =
\frac{1}{\sqrt{N}}\exp\!\left(\iota\frac{2r\pi}{N}j\right)
\end{equation}
with energies
\begin{equation}
\epsilon_r = \xi + 2t_{\perp}\cos\!\left(\frac{2r\pi}{N}\right)
\end{equation}
Taking the limit of $N$ going to infinity one obtains
the gap equation for a fully three dimensional system.
Again assuming only intraband pairing the gap equation
becomes
\begin{equation}
\Delta({\bf q}) = -\sum_{{\bf q}'}
V({\bf q}, {\bf q}')
\frac{\Delta({\bf q}')}{2E({\bf q}')}
\tanh\!\left[\frac{1}{2}\beta E({\bf q}')\right] \\
\label{3d-Gap-eq}
\end{equation}
with the interaction
\begin{equation}
V({\bf q}, {\bf q}') = \frac{1}{N}\left[
V_{\parallel}({\bf q}, {\bf q}')
+ 2\cos(q_z-q_z')V_{\perp}({\bf q}, {\bf q}')\right]  \\
\label{3d-V-def}
\end{equation}
and the usual quasi-particle energies
$E({\bf q})=\sqrt{\epsilon({\bf q})^2+\left|\Delta({\bf
q})\right|^2}$.
In the large $N$ limit this system behaves identically to
the one considered throughout this paper when only singlet
pairing is considered. On the other hand, it is
obvious that this formulation gives very different results
in the small $N$ limit. One might expect that in any
real system the hopping between unit cells is different
from that within a unit cell and so a more general
formulation than either of these would be required.
Such a formulation would, however, yield a continuous
set of fermi levels within some small band and this
result would be in contradiction to experimental
observations.

\section{Interband Pairing}
\label{IB-pairing}

To conclude this discussion we present a more careful
treatment of
the case for which $N=2$ admitting the possibility of pairing
of electrons on different sub-bands of the Fermi-surface.
We will restrict our attention to states
which are even under inversion, neglecting the possibility
of a $p$-wave order parameter.
This necessitates the
consideration of a triplet interlayer pairing state.
The normal state Hamiltonian has the form
\begin{equation}
H_2 = \left(\begin{array}{cc} \xi & -t_{\perp} \\
						-t_{\perp}^{*} & \xi
			\end{array}\right)
\label{2l-H_0}
\end{equation}
where $t_{\perp}=|t_{\perp}|e^{-\iota\varphi}$.
The order parameter has the form
\begin{equation}
\Delta = \left(\begin{array}{cc}
\Delta_{\parallel}^e + \Delta_{\parallel}^o  &
\Delta_{\perp}^s - \iota
\sum_{i=1}^3\Delta_{\perp}^{t,i}\sigma_i  \\
\Delta_{\perp}^s + \iota
\sum_{i=1}^3\Delta_{\perp}^{t,i}\sigma_i  &
\Delta_{\parallel}^e - \Delta_{\parallel}^o
\end{array}\right)
\end{equation}
The $\sigma_i$ are the three Pauli spin matrices.
Upon diagonalizing (\ref{2l-H_0}) we obtain
\begin{eqnarray}
H_2 & = & \left(\begin{array}{cc} \xi - |t_{\perp}| & 0 \\
						0 & \xi + |t_{\perp}|
		\end{array}\right)  \\
\Delta & = & \left(\begin{array}{cc} \Delta_{+} & \Delta_1 -
\iota\Delta_2 \\
						 \Delta_1 + \iota\Delta_2 &
\Delta_{-}
			 \end{array}\right)
\end{eqnarray}
We assume that the triplet component is describable by a
single
complex parameter $\Delta_{\perp}^t$ and a real unit vector
in spin space $\hat{n}$ so that $\Delta_{\perp}^{t,i}=
\Delta_{\perp}^t\hat{n}_i$. The Green's function
can thus be diagonalized in its spin degrees of freedom.
There are, therefore, four independent parameters which
describe the superconducting state. The components of the
order parameter in the band representation are
\begin{equation}
\begin{array}{rcl}
\Delta_{+} & = & \Delta_{\parallel}^e +
\Delta_{\perp}^s \\
\Delta_{-} & = & \Delta_{\parallel}^e -
\Delta_{\perp}^s \\
\Delta_{1} & = & \Delta_{\parallel}^o \\
\Delta_{2} & = &  -\Delta_{\perp}^t\sigma_3
\end{array}
\end{equation}
Designating the bands by $\epsilon_\pm=\xi\mp|t_{\perp}|$
we
solve for the anomalous parts of the Green's function as
before. These may be linearized in $\Delta$ at $T_c$
and upon performing the sum over Matsubara frequencies
we arrive at the following four gap equations
\begin{mathletters}
\begin{eqnarray}
\Delta_+ + \Delta_- & = &
-\sum_{\bf q}V_{\parallel}
\left[\Delta_+
\frac{\tanh\!\left(\frac{1}{2}\beta
\epsilon_{+}\right)}{2\epsilon_{+}}
+ \Delta_-
\frac{\tanh\!\left(\frac{1}{2}\beta \epsilon_{-}\right)}
{2\epsilon_{-}}\right]
\label{Delta-e1-app}
\\
\Delta_+ - \Delta_- & = &
-\sum_{\bf q}V_{\perp}
\left[\Delta_+
\frac{\tanh\!\left(\frac{1}{2}\beta
\epsilon_{+}\right)}{2\epsilon_{+}} - \Delta_-
\frac{\tanh\!\left(\frac{1}{2}\beta \epsilon_{-}\right)}
{2\epsilon_{-}}\right]
\label{Delta-e2-app}
\\
\Delta_1 & = & -\sum_{\bf q}V_{\parallel}
\Delta_1\frac{\tanh\!\left(\frac{1}{2}\beta
\epsilon_{+}\right) +
\tanh\!\left(\frac{1}{2}\beta\epsilon_{-}\right)}
{2\epsilon_{+} + 2\epsilon_{-}}
\label{Delta-o-app}
\\
\Delta_2 & = & -\sum_{\bf q}V_{\perp}
\Delta_2\frac{\tanh\!\left(\frac{1}{2}\beta
\epsilon_{+}\right) +
\tanh\!\left(\frac{1}{2}\beta\epsilon_{-}\right)}
{2\epsilon_{+} + 2\epsilon_{-}}
\label{Delta-t-app}
\end{eqnarray}
\end{mathletters}
If $t_{\perp}=0$ then there is a single transition
temperature associated with the two intralayer pairing
states and another transition temperature associated
with the two interlayer states. Denote the larger one
of these by $T_{c,0}$.
It can be shown that for non-zero $t_{\perp}$ the
transitions
described by
Eqs.~(\ref{Delta-o-app}) and (\ref{Delta-t-app}) have a lower $T_c$
than the intraband pairing state transition in Eqs.~(\ref{Delta-e1-app})
and (\ref{Delta-e2-app})
where for small $t_{\perp}$
the new transition of the interband pairing state \cite{fn-when-IBpairing}
with higher
$T_c$ is given in terms of $T_{c,0}$ by
\begin{equation}
\ln\!\left(\frac{T_c}{T_{c,0}}\right) =
-0.2123\left(\frac{t_{\perp}}{k_BT_c}\right)^2
\end{equation}
The transition temperature for the intraband pairing states
is given by
\begin{equation}
\ln\!\left(\frac{T_c}{T_{c,0}}\right) =
\frac{1}{\left|\ln\frac{T_{c,\parallel}}
{T_{c,\perp}}\right|}
\left(\frac{t_{\perp}R}{2k_BT_c}\right)^2
\end{equation}
where the overlap between the two pair wavefunctions is
\begin{equation}
R = \int\!\!\frac{\mbox{dS}}{(2\pi)^2v_F}\,\psi_{\parallel}\,\psi_{\perp}
\end{equation}
We thus see that in the presence of interlayer tunneling
the favoured pairing state is always an intraband
pairing state.

\begin{table}[tb]
\begin{tabular}{crrr}
$\eta_i({\bf q})$ & \multicolumn{3}{c}{$t_i$} \\
		& \hfill LSCO &
\hfill YBCO & \hfill BSCCO \\ \hline
$1$ & 0 & 0 & 0.1305 \\
$\frac{1}{2}\left(\cos q_x + \cos q_y\right)$ &
-1.0 & -0.50 & -0.5951 \\
$\cos q_x \cos q_y$ &
0.2 & 0.15 & 0.1636 \\
$\frac{1}{2}\left(\cos 2q_x + \cos 2q_y\right)$ &
0 & -0.05 & -0.0519 \\
$\frac{1}{2}\left(\cos 2q_x \cos q_y + \cos q_x \cos
2q_y\right)$ &
0 & 0 & -0.1117 \\
$\cos 2q_x \cos 2q_y$ &
0 & 0 & 0.0510 \\
$\frac{1}{2}\left(\cos q_x - \cos q_y\right)$ &
0 & 0.04 & 0 \\
$\sin q_x \sin q_y$ &
0 & 0 & 0.08
\end{tabular}
\caption{Tight binding basis functions and hopping
parameters
(in eV) used in numerical calculations.}
\label{bs-tab}
\end{table}

\begin{figure}
\caption{Phase diagram showing order parameter symmetry in LSCO:
(a) Near van Hove points; (b) Away from van Hove
points.  The two axes represent the wave vectors $Q_x$,
$Q_y$ at which the pairing interaction has a maximum. }
\label{phase-LSCO-fig}
\end{figure}

\begin{figure}
\caption{Phase diagram for one layer model of YBCO: (a) Near van Hove points;
(b) Away from van Hove
points. The dotted line separates  $A_{1g}$ representations  with and without
$a,b$-axis $\pi$ phase shifts.}
\label{phase-YBCO-fig}
\end{figure}

\begin{figure}
\caption{Phase diagram for one layer model of  BSCCO: (a) Near van Hove points;
(b) Away from van
Hove points. A dotted line separates states with four and eight lobes
within a representation.}
\label{phase-BSCCO-fig}
\end{figure}

\begin{figure}
\caption{ Phase diagram for stability of $\pi$ phase shifts on the
$a$ relative to the $b$ axis, in the order parameter (below solid line)  Model
is for
YBCO with an attractive interlayer and repulsive intralayer
interaction, both peaked at ${\bf Q}_{\scriptbox{AF}}$.
Hole doping fraction appears on the ordinate  and
relative strength of the two interactions defines the abscissa.
The shaded regions denote states where the order parameter on
one (light) or both (dark) bands is nodeless. }
\label{bilayer-fig}
\end{figure}

\begin{figure}
\caption{Evolution of order parameter solutions as interlayer correlations are
increased. The states correspond to the solutions obtained
in the previous figure for $x=0.25$ and the indicated
$\lambda_{\perp}
/ \lambda_{\parallel}$. Observe that proximity to the van Hove
singularity
results in considerable $a,b$-axis anisotropy despite the very small
orthorhombicity.}
\label{s,d-mix-fig}
\end{figure}

\begin{figure}
\caption{Schematic illustration of Josephson coupling between the anti-bonding
band in YBCO (left) and  a Pb counterelectrode (right). Overlap of the single
particle
wave functions is nonzero unless the two wavefunctions are
perfectly
aligned in the directions parallel to the plane of the interface.}
\label{tunnel-fig}
\end{figure}

\begin{figure}
\caption{Possible scenarios for order parameter behavior across twin
boundaries showing  $\pi$ phase shifted $(d,d)$-type (a) or $(s,-s)$-type (b
and c)
solutions. The scenarios depicted in (a) and (c) give $\pi$ junction
behaviour in $a,b$-axis corner junctions while case (b) will
lead to cancellation of the $\pi$ phase shift after averaging
over twin domains.}
\label{twin-fig}
\end{figure}

\begin{figure}
\caption{Amplitude of pure intralayer and interlayer
pairing solutions in the band representation
for $N=4$ (a and b respectively ) and $N=7$ (c and d respectively) cases. The
ideal solutions
($t_{\perp}=0$) are denoted by the dotted lines and the
actual solutions are calculated for moderately large $t_{\perp}$. }
\label{nlayer-fig}
\end{figure}

\begin{figure}
\caption{Solution for a three band problem illustrating co-existence
of solutions of different symmetries.  The inset plots the associated band
structure. }
\label{3l-fig}
\end{figure}

\begin{figure}
\caption{Pairing Density of States (PDOS) of the solution with
highest
$T_c$ in each irreducible representation  for LSCO. Panel (a)
shows the absolute PDOS in arbitrary units and panel (b) illustrates
the PDOS normalized by the single particle density of states. The
isotropic
$s$-state (solid line) was calculated for an attractive pair interaction
peaked at $(\pi,\pi)$ while the other solutions were calculated
for repulsive interactions,  with ${\bf Q}$  chosen so as to maximize
$T_c$.
}
\label{PDOS-fig}
\end{figure}


\begin{thebibliography}{10}
\bibitem{fn-rhombo}
In BSCCO, superlattice effects result in two non-equivalent Cu sites. Thus the
  unit cell is face-centred orthorhombic and double the size of the YBCO unit
  cell with the orthorhombic axes rotated by $45^{\circ}$. For simplicity, we
  simulate this lattice by a unit cell which has ``rhombohedral'' symmetry,
  with a unit cell equivalent to that in YBCO, but with rotated axes of
  symmetry.
\bibitem{fn-sd-mix}
{Since the system is assumed orthorhombic all states are $s$-$d$ mixed states;
  however, the order parameter on a given band will have same (opposite) sign
  unde $\pi /2$ rotation if the $s$ ($d$) component dominates}.
\bibitem{fn-ref-to-fig}
{This effect is illustrated in Figure~\ref{s,d-mix-fig}}.
\bibitem{fn-LGterms}
{It should be noted that there are two more general coupling terms which could,
  in principle, be added to the quartic contribution \cite{Sahu,SigRice}.
  terms do not appear in mean field theory when only pairing of electrons in a
  single band are considered. The analogous terms in the layer language however
  are present}.
\bibitem{fn-show-gamma'}
{When $t_{\perp}$ is small $\gamma'$ is $\frac{0.8525}{(k_BT_c)^2}$ where $T_c$
  is the larger of the two transition temperatures and $\beta'=4\gamma'$.}
\bibitem{fn-delta=0}
{If $\delta=0$ then the sign is undetermined.}
\bibitem{fn-Tc-sup-1}
Calculating the suppression {$T_c$} yields a result similar to that of
  Kettemann and Efetov \cite{KetEf}.
\bibitem{fn-Tc-sup-2}
Finding the suppression of {$T_c$} gives a straightforward extension of the
  usual result \cite{RadLevin-dirt}. Furthermore one finds that if the impurity
  preferentially scatters electrons on one band, then the pairing amplitude is
  increased on the other band.
\bibitem{fn-when-IBpairing}
{Note that for interband pairing to occur we require $|t_{\perp}| < \omega_c$
  since only pairing between time reversed states is being considered.}
\end{thebibliography}
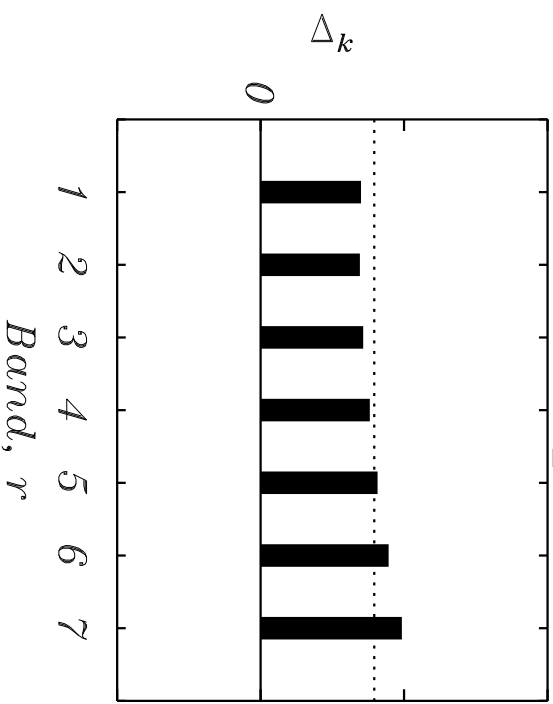
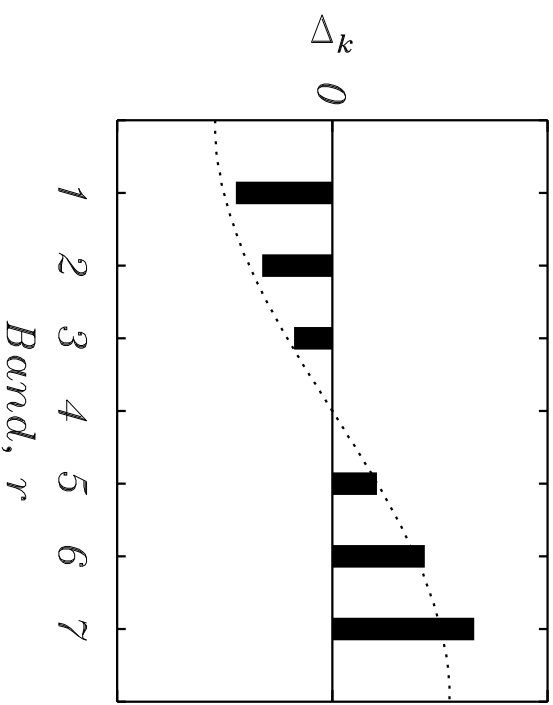
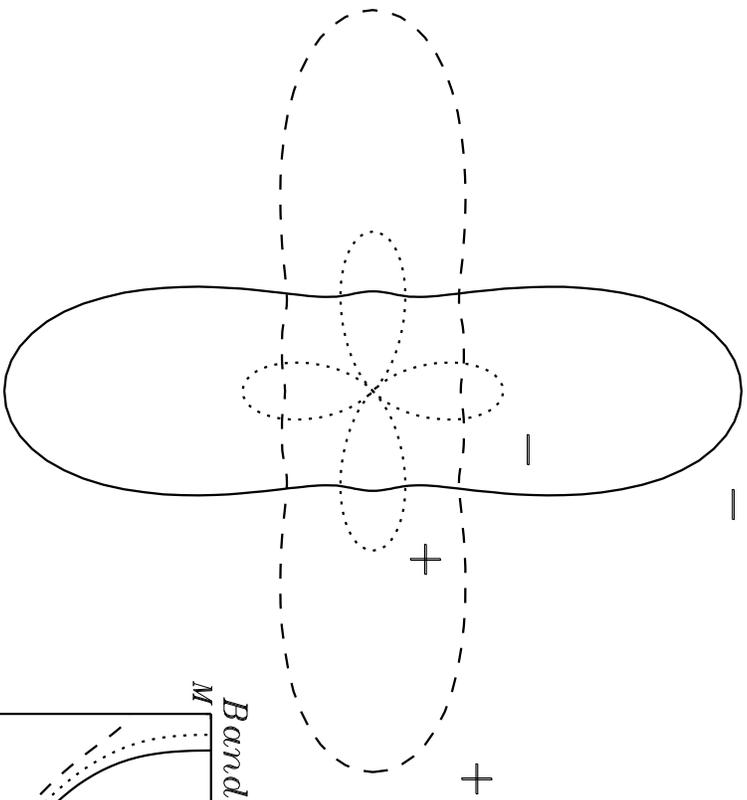
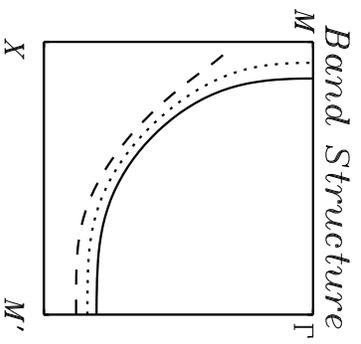
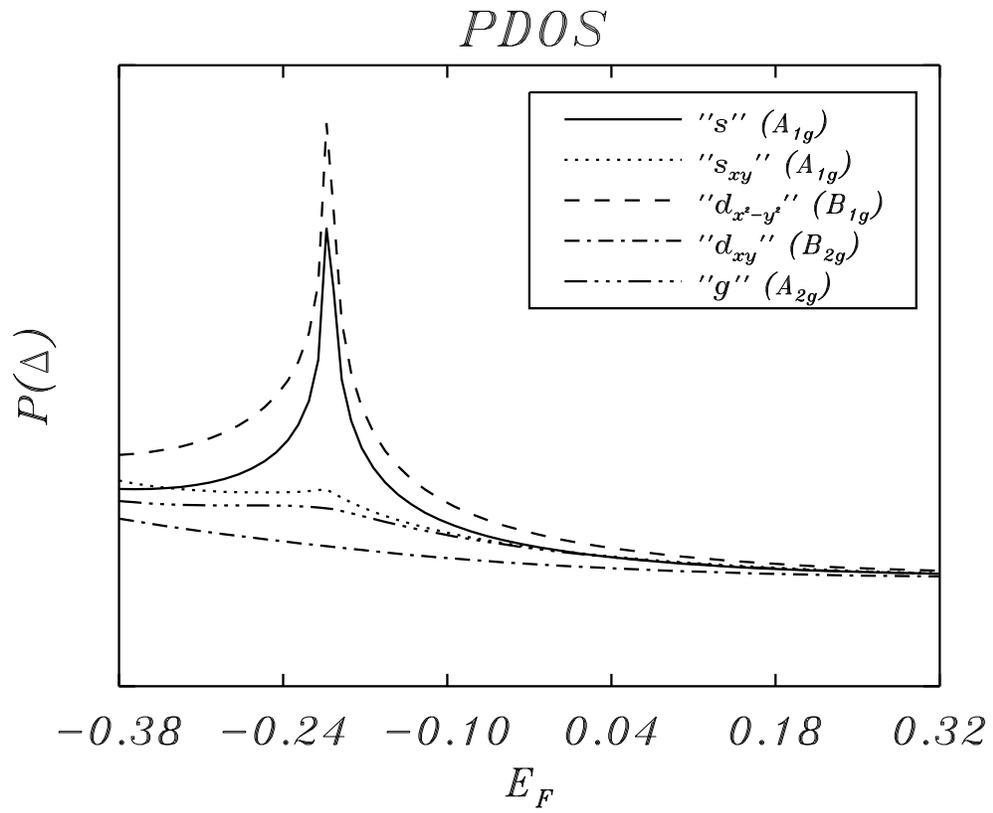
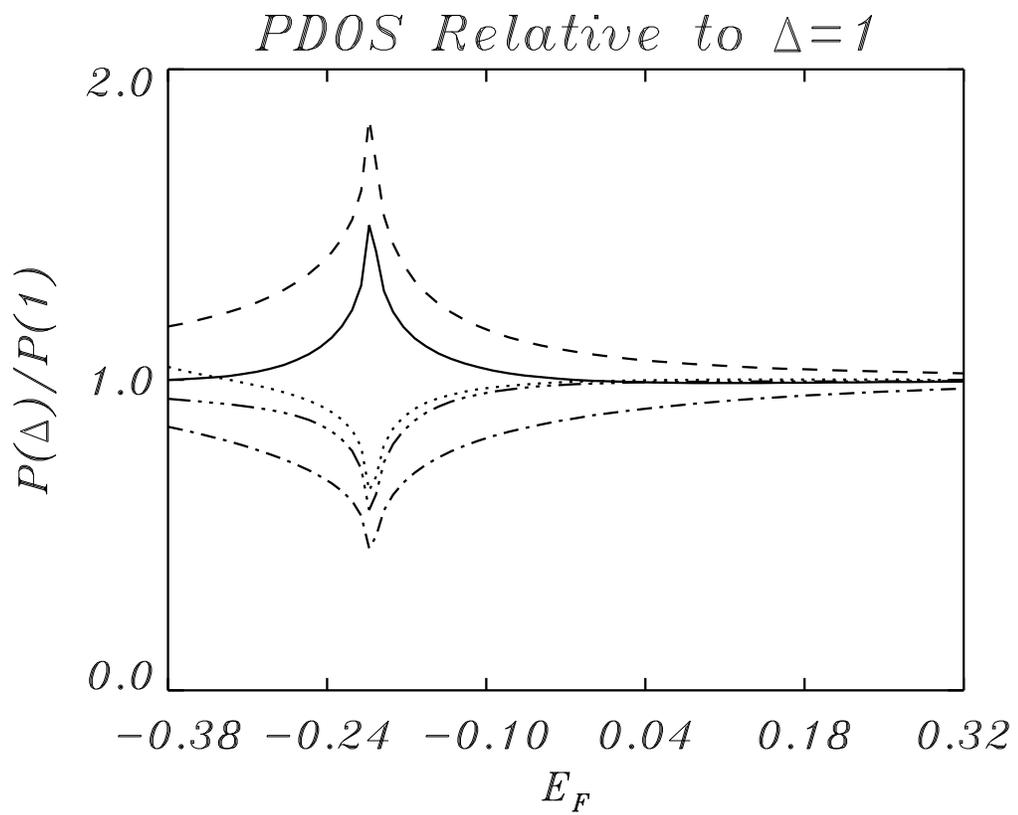
\end{document}